\input psfig.tex

\newcommand {\Mpc}   {\mbox{h$^{-1}$ Mpc \,}}
\newcommand {\kpc}   {\mbox{h$^{-1}$ kpc \,}}
\newcommand {\ks}    {\mbox{km~s$^{-1}$\,}}
\newcommand {\vm}    {\mbox{$<$v$> \,$}}
\newcommand {\vme}   {\mbox{$<$v$>_{ELG}$ \,}}
\newcommand {\vma}   {\mbox{$<$v$>_{non-ELG}$ \,}}
\newcommand {\vn}    {\mbox{$\Delta{v_{n}}$ \,}}
\newcommand {\sv}    {\mbox{$\sigma_{\mbox{{\small v}}}$\,}}
\newcommand {\sve}   {\mbox{$\sigma_{\mbox{\small v},ELG}$ \,}}
\newcommand {\sva}   {\mbox{$\sigma_{\mbox{\small v},non-ELG}$ \,}}
\newcommand {\rh}    {\mbox{r$_{\mbox{\small h}}$ \,}}
\newcommand {\rhe}   {\mbox{r$_{\mbox{\small h},ELG}$ \,}}
\newcommand {\rha}   {\mbox{r$_{\mbox{\small h},non-ELG}$ \,}}

\documentstyle{l-aa}

\begin{document}

\thesaurus{  
       (12.03.3;  
        12.12.1;  
        11.03.1;  
        11.11.1)  
          }


\title{The ESO Nearby Abell Cluster Survey
       \thanks{Based on observations collected at the European Southern
               Observatory (La Silla, Chile)}
      }

\subtitle{III. Distribution and Kinematics of Emission-Line Galaxies}

\author{A.~Biviano \inst{1,5}, P.~Katgert \inst{1}, A.~Mazure \inst{2}, 
        M.~Moles \inst{3,6}, R.~den Hartog \inst{1,7}, J. Perea \inst{3}, 
	P.~Focardi \inst{4}
       }

\institute{Sterrewacht Leiden, The Netherlands \and
   IGRAP, Laboratoire d'Astronomie Spatiale, Marseille, France \and    
   Instituto de Astrof\'{\i}sica de Andaluc\'{\i}a, CSIC, Granada, Spain \and
   Dipartimento di Astronomia, Universit\`a di Bologna, Italy \and
   ISO Science Team, ESA, Villafranca, Spain \and
   Observatorio Astronomico Nacional, Madrid, Spain \and
   ESTEC, SA Division, Noordwijk, The Netherlands
       }

\offprints{P.~Katgert}
\date{Received date; accepted date}

\maketitle
\markboth{The ESO Nearby Abell Cluster Survey: III. The Emission-Line 
          Galaxies}{}

\begin{abstract}

We have used the ESO Nearby Abell Cluster Survey (ENACS) data, to
investigate the frequency of occurrence of Emission-Line Galaxies
(ELG) in clusters, as well as their kinematics and spatial
distribution.

Well over 90\% of the ELG in the ENACS appear to be spirals; however,
we estimate that the detected ELG represent only about one-third of
the total spiral population.

The {\em apparent} fraction of ELG increases towards fainter
magnitude, as redshifts are more easily obtained from emission lines
than from absorption lines. From the ELG that have an absorption-line
redshift as well, we derive a {\em true} ELG fraction in clusters of
0.10, while the {\em apparent} fraction is 0.16.
 
The {\em apparent} ELG fraction in the field is 0.42, while the {\em
true} fraction is 0.21. The {\em true} ELG fractions in field and
clusters are consistent if the differences in morphological mix are
taken into account. Thus, it is not necessary to assume that ELG in
and outside clusters have different emission-line properties.

The average ELG fraction in clusters depends on global velocity
dispersion \sv: the {\em true} fraction decreases from $0.12$ for
$\sv\la 600$ \ks to $0.08$ for $\sv\ga 900$ \ks.

In only 12 out of 57 clusters, the average velocity of the ELG differs
by more than $2 \sigma$ from that of the other galaxies, and in only 3
out of 18 clusters \sv of the ELG differs by more than $2 \sigma$ from
that of the other galaxies. Yet, combining the data for 75 clusters,
we find that \sv of the ELG is, on average, 20 \% larger than that of
the other galaxies. It is unlikely that this is primarily due to
velocity offsets of the ELG w.r.t. the other galaxies; instead, the
larger \sv for the ELG must be largely intrinsic.

The spatial distribution of the ELG is significantly less peaked
towards the centre than that of the other galaxies. This causes the
average projected density around ELG to be $\sim$ 30\% lower than it
is around the other galaxies. In combination with the inevitable
magnitude bias against galaxies without detectable emission lines,
this can lead to serious systematic effects in the study of distant
clusters. 

From an analysis of the distributions of projected pair distances and
velocity differences we conclude that at most 25\% of the ELG are in
compact substructures, while the majority of the ELG are distributed
more or less smoothly.

The virial estimates of the cluster masses based on the {\em ELG only}
are, on average, about 50\% higher than those derived from the other
galaxies. This indicates that the ELG are either on orbits that are
significantly different from those of the other galaxies, or that the
ELG are not in virial equilibrium with the other galaxies, or both.

The velocity dispersion profile of the ELG is found to be consistent
with the ELG being on more radial orbits than the other galaxies. For
the ELG, a ratio between tangential and radial velocity dispersion of
0.3 to 0.8 seems most likely, while for the other galaxies the data
are consistent with isotropic orbits.

The lower amount of central concentration, the larger value of \sv and
the possible orbital anisotropy of the ELG, as well as their content
of line-emitting gas would be consistent with a picture in which
possibly all spirals (but certainly the late-type ones) have not yet
traversed the virialized cluster core, and may even be on a first
(infall) approach towards the central, high-density region.

\end{abstract}


\section{Introduction}
\label{s-intro}

Galaxies of different morphological types live in different
environments (e.g. Hubble \& Humason 1931). Dressler (1980a) was the
first to clearly establish the dependence of the fractions of early-
and late-type cluster galaxies on the local galaxy density. The
dependence found by Dressler has been verified by many authors, most
recently by e.g. Binggeli, Tarenghi \& Sandage (1990); Sanrom\`a \&
Salvador-Sol\'e (1990); Iovino et al. (1993).

Early and late-type cluster galaxies not only differ in their spatial
distribution, but also in their kinematics. Moss \& Dickens (1977)
claimed that the velocity dispersion, \sv, of the population of
late-type galaxies is significantly larger than that of the early-type
galaxies, in 4 of the 5 clusters for which they could determine
velocity dispersions for early- and late-type galaxies separately.
Their study was a follow-up of earlier suggestions that the kinematics
of early- and late-type galaxies in the Virgo cluster are different.
Differences in average velocity, \vm (de Vaucouleurs 1961), as well as
in \sv (Tammann 1972) had been reported. Only the \sv-difference was
subsequently confirmed (Binggeli, Tammann, \& Sandage 1987). The early
claim of Moss \& Dickens (1977) was confirmed by Sodr\'e et al. (1989)
and Biviano et al. (1992), from data on galaxies in 15 and 37 galaxy
clusters respectively.

In clusters, the dependence of the mix of morphological types on local
density (i.e.\ on distance from the cluster center), and the
differences in kinematics that are related to this, can generally be
understood as the result of the evolution of the galaxy population.
Several processes may affect the morphology of a galaxy as it passes
through the dense cluster core (e.g. ram pressure, merging, tidal
stripping and tidal shaking). These processes are believed to be
capable of transforming a star-forming spiral galaxy in a quiescent
elliptical or S0. On the other hand, it is possible that regions of
high density are, from the start, more conducive to the formation of
slowly spinning (early-type~?) galaxies (see e.g.  Sarazin 1986, and
reference therein). It is likely that clusters form mainly through the
collapse of density perturbations (e.g. Gunn \& Gott 1972) although it
is possible that shear also plays a r\^ole. If such density
perturbations have density profiles that fall with radius, it is
natural to expect a time sequence of infalling shells of galaxies. The
spirals could then be on infalling orbits, as was convincingly shown
to be the case in the Virgo cluster by Tully \& Shaya (1984), whereas
the ellipticals and S0's would constitute the virialized cluster
population.

Some recent findings indicate that the latter scenario may well be too
simplistic. On the one hand, Zabludoff \& Franx (1993) have found that
the early- and late-type galaxies have different average velocities in
three out of six clusters studied, while the \sv 's are not
different. On the other hand, Andreon (1994) carefully re-examined
galaxy morphologies in the Perseus cluster, and did not find a clear
morphology-density relation. If groups of galaxies fall into a cluster
anisotropically (as suggested e.g.\ by van Haarlem \& van de Weygaert
1993), this may result in an average velocity of the infalling
(spiral?) population that differs from that of the other galaxies in
the (core of the) cluster. The resulting substructure could, at the
same time, wash out the morphology-density relation.

Previous investigations of emission-line galaxies (ELG) in and outside
clusters (Gisler 1978; Dressler, Thompson \& Shectman 1985; Salzer et
al. 1989; Hill \& Oegerle 1993; Salzer et al. 1995) have been mainly
limited to the comparison of the relative frequency of ELG in clusters
and in the field. These studies have shown that emission lines occur
more frequently in the spectra of field galaxies than in cluster
galaxies (for elliptical galaxies this was already pointed out by
Osterbrock, 1960). It was concluded that this difference cannot
totally be the result of the morphology-density relation, in
combination with the different mix of early- and late-type
galaxies. However, recently the kinematics of the ELG has become a
subject of study (e.g. Mohr et al. 1996; Carlberg et al. 1996).

In this paper, we re-examine the evidence for differences between
early- and late-type galaxies in clusters, by using the extensive
data-base provided by the ENACS (the ESO Nearby Abell Cluster
Survey). We analyze the frequency of occurrence of ELG in clusters, as
well as their distribution with respect to velocity and position and
their kinematics. In \S~\ref{s-data} we summarize those properties of
the ENACS data-base that are relevant for the present discussion. In
\S~\ref{s-frac} we discuss the fraction of ELG in clusters and in the
field. In \S~\ref{s-kine} and \S~\ref{s-spatial} we study the global
kinematics and spatial distribution of the ELG in relation to the
non-ELG. In \S~\ref{s-substr} we discuss correlations between
positions and velocities of the ELG and non-ELG. In \S~\ref{s-formev}
we investigate the equilibrium and the orbits of the cluster galaxies,
and, finally, in \S~\ref{s-discuss} we discuss the implications of our
results for ideas about structure and formation of clusters.

\section{The Data}
\label{s-data}
\subsection{The ESO Nearby Abell Cluster Survey}
\label{ss-enacs}

The ENACS has provided reliable redshifts for 5634 galaxies in the
directions of 107 cluster candidates from the catalogue of Abell,
Corwin \& Olowin (1989), with richness $R_{\rm ACO} \geq 1$ and mean
redshift $z \la 0.1$. Redshift estimates are mostly based on
absorption lines, but for 1231 galaxies, emission lines were
detectable in the spectrum.  As described in Katgert et al.\ (1996,
hereafter Paper I), for 62 galaxies the reality of the emission lines
is doubtful, as judged from a comparison with the absorption-line
redshift (in almost all cases these are galaxies with only one
emission line detected in the spectrum). That leaves 1169 galaxies
with reliable emission lines. For 586 of these, the redshift is based
on both absorption and emission lines, and for the remaining 583
galaxies the redshift estimate is based exclusively on emission
lines. The estimated redshift errors range from about 40 to slightly
over 100 km/s, with the majority less than 70 km/s. For a detailed
description of the characteristics of the ENACS data-base we refer to
Paper I.

For almost all of the 5634 galaxies a calibrated R-magnitude estimate
is available, from photographic photometry calibrated with
CCD-imaging. The R-magnitudes of the galaxies with redshifts range
from 13 to about 18, although the majority of the galaxies have
R-magnitudes brighter than about 17.

For most of the galaxies in this survey we could not obtain a reliable
morphological classification because the galaxies were identified on
copies of survey plates made with Schmidt telescopes. However, we can
identify star-forming (i.e. presumably late-type) galaxies on the
basis of the presence of the relevant emission lines in their spectra.
The clear {\em advantage} of selecting galaxies on the presence of
spectral lines is that the selection is quite effective out to
redshifts of $z=0.1$, whereas it may already be difficult to reliably
determine morphologies of galaxies at a redshift $z \approx 0.05$
(e.g.  Andreon 1993). The {\em disadvantage} of a selection on the
basis of detectable emission-lines is that the absence of such lines
does not decisively prove a galaxy to be an early-type galaxy. In
other words: the class of galaxies without detectable emission lines
is likely to contain also late-type galaxies with emission lines that
are too faint to be detected, or without emission lines. In the
following, we will nevertheless refer to the two galaxy populations as
ELG and non-ELG. The ELG can be thought of as an almost pure late-type
galaxy population (see \S~\ref{ss-clusfld}), whereas the non-ELG are a
mix of early- and late-type galaxies (which share the property that
they do not have detectable emission lines).

\subsection{The Definition of Redshift Systems}
\label{ss-systems}

The 107 pencil beam redshift surveys cover solid angles with angular
diameters between 0.5 and about 1.0 $\deg$. In these 107 redshift
surveys, 220 systems were found that are compact in redshift and that
contain at least 4 (but often several tens to a few hundred) member
galaxies. These systems were identified in redshift space, by using
the method of fixed gaps (see Paper I), which separates galaxies
within a system (with velocity differences between `neighbours' less
than the chosen gap) from galaxies that do not belong to the system
(because the velocity difference with the nearest system member is
larger than the chosen gap). For the following discussion we will, in
addition, divide the cluster Abell~548 into two components, following
the suggestion of Escalera et al. (1994) (which was later confirmed by
Davis et al. 1995) because the cluster is clearly bi-modal.

Membership of a given galaxy to a particular system requires that the
galaxy has a velocity within the velocity limits of the system as
defined with the fixed-gap method. For systems with at least 50
galaxies we applied an additional test for membership which uses both
the velocity {\em and} position (see den~Hartog \& Katgert 1996; see
also Mazure et al. 1996, hereafter Paper II). This second criterion
removes 74 galaxies for which the combination of position in the
cluster and relative radial velocity makes it unlikely that they are
within the turn-around radius of their host system. These 74
`interlopers' occur in only 25 of the systems.

The `interloper'-test involves an estimate of the mass-profile of the
system, and therefore requires the centre of the system. Following
den~Hartog \& Katgert (1996), we have assumed the centre to be (in
order of preference): 1) the X-ray center, 2) the position of the
brightest cluster member in the cluster core, provided it is at least
one magnitude brighter than the second brightest member, and/or less
than 0.25~\Mpc from the geometric center of the galaxy distribution.
If these two methods could not be applied, we determined 3) the
position of the peak in the surface density, viz. the position of the
galaxy with the smallest distance to its N$^{1/2}$-th neighbour (with
N the number of galaxies in the system). In several cases this
position differed by more than 0.1~\Mpc from that of any of the 3
brightest cluster members. We then used 4) a luminosity weighted
average position. If the latter was not nearer than 0.25~\Mpc to the
geometric center, we used 5) the geometric center, as defined by the
{\em biweight} averages of the galaxy positions (see, e.g., Beers,
Flynn, \& Gebhardt 1990). For 22 of the 25 systems with at least 50
galaxies, the position of the X-ray peak or that of the brightest
cluster member were chosen as cluster centers.

\subsection{The Various Samples of Galaxy Systems}
\label{ss-samples}

Our discussion of the differences between the average velocities of
ELG and non-ELG within individual clusters will only be based on the
58 systems that contain at least 5 ELG: we consider this a minimum
number for the estimation of a meaningful average velocity. In
general, such systems also contain at least 5 non-ELG. However, for
A3128 ($z~=~0.077$), the number of non-ELG is less than 5 and we have
therefore not considered it in the analysis for individual systems.
That leaves a sample of 57 systems (sample 1) with both at least 5 ELG
{\em and} 5 non-ELG.

In discussing velocity dispersions of {\em individual} systems we have
limited ourselves to the subset of 18 systems with at least 10 ELG
(all of which also have at least 10 non-ELG). I.e.\ we applied a lower
limit to the ELG population that is identical to the one used in Paper
II, in the discussion of the distribution of velocity dispersions of a
complete volume-limited sample of rich clusters. The same restriction
was applied in estimating projected harmonic mean radii: from
numerical modeling we find that such estimates are biased if they
are based on less than 10 positions.  The sample of 18 systems with at
least 10 ELG will be referred to as sample 2.

Finally, we will also discuss results for a sample of 75 systems with
at least 20 members (sample 3). The requirement that the total number
of galaxies in a system be at least 20 ensures that the centre of the
system can be determined with sufficient accuracy. This sample also
defines a `synthetic' average cluster, which contains 3729 galaxies of
which 559 are ELG.

\begin{table}
\caption[]{The data-set of 120 systems}
\begin{flushleft}
\small
\begin{tabular}{rrcrrc}
\hline
\noalign{\smallskip}
Abell & $<z>$ & Center & \multicolumn{2}{c}{$N,N_{ELG}$} & r$_{max}$ \\
\noalign{\smallskip}
& & $\alpha$~~~~~~~~$\delta$ & & & Mpc \\
& &    B1950        & &  \\
\hline
\noalign{\smallskip}
  13\ \ \ \ &0.094 & 00:11:02 --19:45.7& 37& 3 &1.6\\
  87\ \ \ \ &0.055 & 00:40:13 --10:04.3& 27& 2 &0.6\\
 118\ \ \ \ &0.115 & 00:52:52 --26:37.3& 30& 8 &1.3\\
 119\ \ \ \ &0.044 & 00:53:45 --01:31.6&101& 5 &1.2\\
 151\ \ \ \ &0.041 & 01:06:27 --16:12.7& 25& 5 &0.8\\
 151\ \ \ \ &0.053 & 01:06:22 --15:40.4& 46&10 &1.6\\
 151\ \ \ \ &0.099 & 01:06:08 --15:53.3& 35& 5 &2.0\\
 168\ \ \ \ &0.045 & 01:12:35 +00:01.4& 76& 6 &1.1\\
 229\ \ \ \ &0.113 & 01:36:44 --03:53.1& 32& 8 &1.3\\
 295\ \ \ \ &0.043 & 01:59:44 --01:22.1& 30& 1 &0.5\\
 367\ \ \ \ &0.091 & 02:34:18 --19:35.2& 27& 3 &1.1\\
 380\ \ \ \ &0.134 & 02:41:60 --26:26.3& 25& 4 &1.8\\
 420\ \ \ \ &0.086 & 03:06:56 --11:46.8& 19& 5 &1.2\\
 514\ \ \ \ &0.071 & 04:46:21 --20:37.3& 81&11 &1.7\\
 524\ \ \ \ &0.056 & 04:55:42 --19:45.1& 10& 2 &0.7\\
 524\ \ \ \ &0.078 & 04:55:40 --19:47.0& 26&12 &0.9\\
 543\ \ \ \ &0.085 & 05:29:19 --22:19.8& 10& 1 &0.7\\
 548W       &0.042 & 05:43:34 --25:53.3&120&24 &1.6\\
 548E \     &0.041 & 05:46:38 --25:29.3&114&38 &1.5\\
 548\ \ \ \ &0.087 & 05:43:36 --25:28.8& 14& 8 &4.6\\
 548\ \ \ \ &0.101 & 05:43:47 --25:42.4& 21& 6 &3.1\\
 754\ \ \ \ &0.055 & 09:06:49 --09:28.8& 39& 0 &0.8\\
 957\ \ \ \ &0.045 & 10:11:08 --00:41.3& 34& 1 &0.6\\
 978\ \ \ \ &0.054 & 10:17:56 --06:16.5& 61& 7 &1.5\\
1069\ \ \ \ &0.065 & 10:37:14 --08:25.8& 35& 0 &0.8\\
1809\ \ \ \ &0.080 & 13:50:36 +05:23.6& 30& 0 &0.9\\
2040\ \ \ \ &0.046 & 15:10:21 +07:36.7& 37& 3 &0.6\\
2048\ \ \ \ &0.097 & 15:12:50 +04:34.0& 25& 1 &1.2\\
2052\ \ \ \ &0.035 & 15:14:18 +07:12.4& 35& 2 &0.4\\
2353\ \ \ \ &0.121 & 21:31:47 --01:47.9& 24& 4 &1.4\\
2361\ \ \ \ &0.061 & 21:36:08 --14:32.3& 24& 7 &0.9\\
2362\ \ \ \ &0.061 & 21:37:31 --14:27.5& 17& 5 &1.1\\
2401\ \ \ \ &0.057 & 21:55:36 --20:20.6& 23& 1 &0.6\\
2426\ \ \ \ &0.088 & 22:11:19 --10:24.0& 11& 0 &1.4\\
2426\ \ \ \ &0.098 & 22:11:52 --10:37.4& 15& 1 &1.0\\
2436\ \ \ \ &0.091 & 22:17:59 --03:04.9& 14& 0 &1.1\\
2480\ \ \ \ &0.072 & 22:43:18 --17:53.3& 11& 1 &0.8\\
2500\ \ \ \ &0.078 & 22:50:48 --25:49.3& 12& 6 &0.8\\
2500\ \ \ \ &0.090 & 22:51:03 --25:46.0& 13& 4 &0.8\\
2569\ \ \ \ &0.081 & 23:14:54 --13:05.7& 36& 2 &1.2\\
2644\ \ \ \ &0.069 & 23:38:18 --00:11.1& 12& 0 &1.2\\
2715\ \ \ \ &0.114 & 00:00:12 --34:57.3& 14& 1 &1.3\\
2717\ \ \ \ &0.049 & 24:00:40 --36:12.9& 40& 2 &1.3\\
2734\ \ \ \ &0.062 & 00:08:50 --29:07.9& 77& 1 &1.7\\
2755\ \ \ \ &0.095 & 00:15:11 --35:28.7& 22& 3 &1.2\\
2755\ \ \ \ &0.121 & 00:16:19 --35:25.4& 10& 2 &1.6\\
2764\ \ \ \ &0.071 & 00:18:08 --49:29.4& 19& 3 &1.0\\
2765\ \ \ \ &0.080 & 00:19:01 --21:02.1& 16& 9 &0.9\\
2778\ \ \ \ &0.102 & 00:26:25 --30:26.6& 17& 9 &1.6\\
2778\ \ \ \ &0.119 & 00:25:22 --30:33.7& 10& 5 &1.5\\
2799\ \ \ \ &0.063 & 00:35:02 --39:24.3& 36& 5 &0.8\\
2800\ \ \ \ &0.064 & 00:35:29 --25:20.9& 34& 6 &1.0\\
\hline	   
\normalsize
\end{tabular}
\end{flushleft}
\label{t-data1}
\end{table}

\begin{table}
\addtocounter{table}{-1}
\caption[]{Continued}
\begin{flushleft}
\small
\begin{tabular}{rrcrrc}
\hline
\noalign{\smallskip}
Abell & $<z>$ & Center & \multicolumn{2}{c}{$N,N_{ELG}$} & r$_{max}$ \\
\noalign{\smallskip}
& & $\alpha$~~~~~~~~$\delta$ & & & Mpc \\
& &    B1950        & &  \\
\hline
\noalign{\smallskip}
2819&0.075 & 00:43:46 --63:49.0& 49& 6 &1.4\\
2819&0.087 & 00:43:54 --63:52.2& 43& 6 &1.9\\
2819&0.160 & 00:41:46 --64:11.8& 13& 1 &4.8\\
2854&0.061 & 00:58:34 --50:48.2& 22& 4 &0.8\\
2871&0.114 & 01:05:52 --37:01.6& 14& 3 &1.2\\
2871&0.123 & 01:05:31 --36:59.4& 18& 4 &1.4\\
2911&0.081 & 01:23:51 --38:13.5& 31& 2 &1.0\\
2923&0.072 & 01:30:03 --31:20.9& 16& 3 &0.8\\
3009&0.065 & 02:20:17 --48:47.5& 12& 3 &0.8\\
3093&0.083 & 03:09:15 --47:35.1& 22& 5 &0.9\\
3094&0.067 & 03:09:49 --27:09.9& 66&16 &1.5\\
3094&0.139 & 03:09:19 --27:16.9& 12& 1 &3.2\\
3111&0.078 & 03:15:55 --45:51.8& 35& 3 &1.0\\
3112&0.075 & 03:16:13 --44:24.9& 67&16 &1.5\\
3112&0.132 & 03:16:15 --44:26.8& 14& 1 &2.3\\
3122&0.064 & 03:20:21 --41:31.4& 89&18 &1.7\\
3122&0.150 & 03:20:01 --42:03.5& 10& 2 &2.1\\
3128&0.039 & 03:29:52 --52:36.3& 12& 1 &1.5\\
3128&0.060 & 03:29:27 --52:40.7&152&30 &2.4\\
3128&0.077 & 03:28:27 --53:13.9& 11& 7 &3.0\\
3141&0.105 & 03:34:55 --28:11.0& 15& 0 &1.2\\
3142&0.066 & 03:34:35 --39:53.3& 12& 3 &1.1\\
3142&0.103 & 03:34:56 --39:57.9& 21& 2 &1.1\\
3151&0.068 & 03:38:22 --28:50.2& 38& 6 &0.8\\
3158&0.059 & 03:41:38 --53:47.5&105& 9 &1.7\\
3194&0.097 & 03:57:11 --30:18.7& 32& 8 &1.3\\
3202&0.069 & 03:59:24 --53:49.3& 27& 4 &0.9\\
3223&0.060 & 04:06:34 --30:57.2& 73& 6 &1.5\\
3341&0.038 & 05:23:40 --31:35.0& 63&11 &0.8\\
3341&0.078 & 05:22:32 --31:39.6& 15& 4 &1.7\\
3341&0.115 & 05:21:42 --31:43.0& 18& 1 &4.0\\
3354&0.059 & 05:33:04 --28:34.2& 57&10 &1.5\\
3365&0.093 & 05:46:14 --21:56.5& 32& 5 &1.0\\
3528&0.054 & 12:51:41 --28:45.2& 28& 0 &1.1\\
3558&0.048 & 13:25:08 --31:14.3& 73& 9 &1.5\\
3559&0.047 & 13:27:04 --29:15.4& 39&10 &1.6\\
3559&0.113 & 13:24:59 --29:08.4& 11& 1 &2.9\\
3562&0.048 & 13:30:48 --31:24.9&116&21 &2.2\\
3651&0.060 & 19:48:10 --55:11.4& 78& 8 &1.9\\
3667&0.056 & 20:08:27 --56:58.6&103& 9 &1.8\\
3682&0.092 & 20:25:59 --37:07.8& 10& 1 &0.9\\
3691&0.087 & 20:30:55 --38:12.7& 33& 2 &1.0\\
3693&0.091 & 20:31:15 --34:48.1& 16& 5 &1.2\\
3695&0.089 & 20:31:33 --35:59.4& 81& 9 &1.9\\
3696&0.088 & 20:32:23 --35:09.8& 12& 0 &1.2\\
3703&0.073 & 20:35:53 --61:30.7& 18& 5 &0.9\\
3703&0.091 & 20:35:59 --61:25.7& 13& 2 &1.3\\
3705&0.090 & 20:38:54 --35:23.9& 29& 3 &1.0\\
3733&0.039 & 20:59:01 --28:15.4& 41& 6 &0.6\\
3744&0.038 & 21:04:30 --25:37.8& 66&13 &1.1\\
3764&0.076 & 21:22:48 --34:56.9& 38&10 &0.9\\
3795&0.089 & 21:35:54 --32:17.9& 13& 3 &1.3\\
\hline
\normalsize
\end{tabular}
\end{flushleft}
\end{table}

\begin{table}
\addtocounter{table}{-1}
\caption[]{Continued}
\begin{flushleft}
\small
\begin{tabular}{rrcrrc}
\hline
\noalign{\smallskip}
Abell & $<z>$ & Center & \multicolumn{2}{c}{$N,N_{ELG}$} & r$_{max}$ \\
\noalign{\smallskip}
& & $\alpha$~~~~~~~~$\delta$ & & & Mpc \\
& &    B1950        & &  \\
\hline
\noalign{\smallskip}
3799&0.045 & 21:36:36 --72:51.9& 10& 4 &0.6\\
3806&0.076 & 21:42:50 --57:31.0& 97&23 &2.3\\
3809&0.062 & 21:43:49 --44:07.8& 89&21 &1.7\\
3809&0.110 & 21:46:38 --43:58.5& 10& 4 &3.9\\
3809&0.142 & 21:42:20 --44:03.0& 11& 1 &4.5\\
3822&0.076 & 21:50:34 --58:06.2& 84&15 &1.9\\
3825&0.075 & 21:55:06 --60:34.3& 59& 4 &1.7\\
3825&0.104 & 21:53:39 --60:27.5& 17& 7 &1.9\\
3827&0.098 & 21:58:26 --60:10.8& 20& 1 &1.6\\
3864&0.102 & 22:16:58 --52:43.0& 32& 6 &1.2\\
3879&0.067 & 22:24:05 --69:16.7& 45& 9 &1.5\\
3897&0.073 & 22:36:30 --17:36.1& 10& 0 &1.0\\
3921&0.093 & 22:46:41 --64:41.7& 32& 7 &1.4\\
4008&0.055 & 23:27:49 --39:33.5& 27& 3 &0.7\\
4010&0.096 & 23:28:34 --36:47.2& 30& 6 &1.2\\
4053&0.072 & 23:52:11 --27:57.6& 17& 5 &0.7\\
\hline
\normalsize
\end{tabular}
\end{flushleft}
\end{table}

In Tab.~\ref{t-data1} we list some characteristics of all 87 systems
in the 3 samples defined above, as well as of the 33 systems with a
total number of members from 10 to 19, of which less than 5 are ELG.
The total number of galaxies in these 120 systems is 4333, of which
809 are ELG. In col.(1) the ACO number (Abell et al.\ 1989) of the
(parent) ACO system is given, and in col.(2) the average redshift of
the system. Col.(3) gives the position of the centre of the system
(B1950.0). The number of member galaxies, and the number of member ELG
among these are given in col.(4) (note that these numbers do not
include the 74 interlopers), and col.(5) lists the projected distance,
r$_{max}$, in \Mpc, of the galaxy farthest from the cluster centre.

If there are several systems along the line-of-sight to a given
cluster, these are identified by their average redshift, which was
obtained using the biweight estimator. Throughout this paper, averages
are determined with the biweight estimator, since this is
statistically more robust and efficient than the standard mean in
computing the central location of a data-set (see Beers et
al. 1990). When at least 15 velocities are available, velocity
dispersions were also computed with the biweight estimator; however,
for smaller number of redshifts we used the gapper estimator.  These
estimators yield the best robust estimates of the true values of
location and scale of a given data-set, particularly when outliers are
present.

\subsection{The Emission-line Galaxies}
\label{ss-emiss}

In the wavelength range covered by the ENACS observations, and for the
redshifts of the clusters studied, the principal emission lines that
are observable are [OII] (3727 \AA), H$\beta$ (4860 \AA) and the
[OIII] doublet (4959, 5007 \AA). Note that because of the small
aperture of the Optopus fibers (2.3 arcsec diameter), we have only
sampled emission-lines in the very central regions of the galaxies.
For the redshifts of our clusters the diameters of these regions are
$2.5~\pm~0.8$~\kpc. This should be kept in mind when making
comparisons with other datasets for which the information about
emission lines may refer to much larger or smaller apertures.

The emission lines were identified independently by two of us, in two
different ways; first by examining the 2~-~D Optopus CCD frames, and
second by inspecting the uncleaned wavelength-calibrated 1-D spectra.
Two lists of candidate ELG were thus produced, and for the relatively
small number of cases in which there was no agreement, both the 2-D
frames and 1-D spectra were examined again. The inspection of the 2-D
frames allowed easy discrimination against cosmic-ray events (emission
lines are soft and round as they are images of the fiber), and against
sky-lines (since these are found at the same wavelength in all
spectra). While examining the 1-D spectra we also obtained the
wavelengths of the emission lines by fitting Gaussians superposed on
a continuum to them.

The combined list of galaxies that show emission lines contains 1231
ELG. As mentioned earlier, for 62 of these we have good evidence that
the emission line(s) are not real; in the large majority of these
cases there is only one line. For a subset of 586 of the remaining
1169 ELG, the reality of the emission lines is borne out by the very
good agreement between the absorption- and emission-line redshifts
(see Paper I). For the other 583 ELG, no confirmation of the reality
of the lines is available; we expect that in at most 10\% of these
cases the lines are not real.

Among the 1169 ELG there are 78 active galactic nuclei (AGN). These
were identified either through the large velocity-width of the
H$\beta$ line, or through the intensity ratios of the [OIII] and
H$\beta$ lines, and the relative intensity of the [OII] line (if
present). We are convinced that our criteria were sufficiently strict
that all 78 galaxies that we classify as AGN are indeed {\em bona
fide} AGN. However, at the same time, our criteria were probably too
strict to identify all AGN in our dataset.

It should be realized that our ELG sample is not complete w.r.t. a
well-defined limit in equivalent width of the various emission
lines. Furthermore, the poorly-defined limit in equivalent width is
probably not sufficiently low that essentially all galaxies with
emission lines will have been identified as ELG. Therefore, the sample
of non-ELG is very likely to contain a mix of real non-ELG (i.e.
galaxies without emission-lines) and unrecognized ELG with emission
lines that are too weak to be detected in the ENACS observations. Any
difference between ELG and non-ELG that we may detect is therefore
probably a reduced version of a real difference. For the same reason,
the absence of an observable difference between ELG and non-ELG does
not prove conclusively that there is no difference between the ELG and
the other galaxies.

\subsection{Completeness w.r.t. Apparent Magnitude}
\label{ss-compl}

As was discussed in Paper I, spectroscopy was attempted for all
galaxies in the fields of the target ACO clusters down to well-defined
limits in isophotal magnitude. However, the success rate of the
determination of an {\em absorption-line} redshift depends strongly on
the signal-to-noise ratio in the galaxy spectrum. This in turn depends
primarily on the surface brightness of that part of the galaxy that
illuminates the fibre entrance. As a result, the success rate is
highest for intermediate magnitudes, and decreases somewhat for
brighter galaxies (as those are large, so that only a small fraction
of the total flux is sampled), and quite noticeably for fainter
galaxies for which the {\em total} flux is smaller. On the contrary,
the succes rate for the detection of emission lines does not appear to
depend significantly on the brightness of the galaxy.  Therefore, the
relative distribution w.r.t.\ magnitude of ELG and non-ELG can be
different, as it is generally easier to obtain a redshift for a faint
galaxy if it has emission-lines in its spectrum, than if it has not.

\begin{figure}
\psfig{file=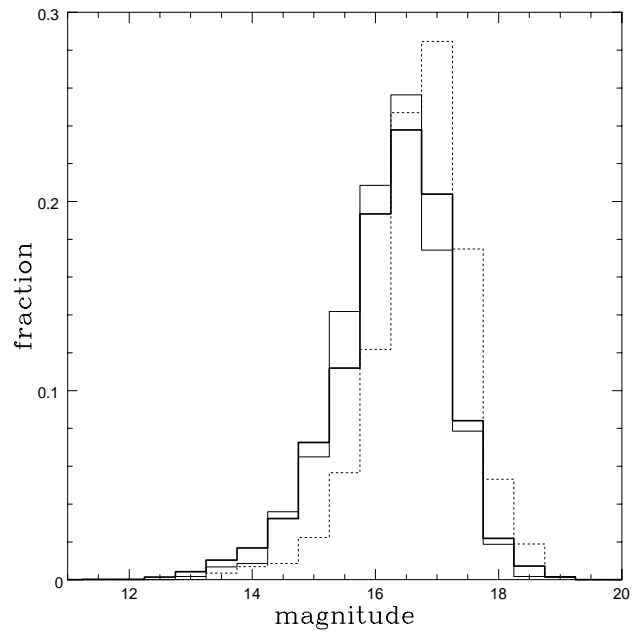,height=9.0cm,width=8.8cm}
\caption[]{The normalized distribution w.r.t. apparent magnitude 
(R$_{25}$) for three subsets of the ENACS: the 4447 galaxies with
redshift based solely on absorption lines (heavy-line histogram), the
585 galaxies with redshift based both on absorption and emission lines
(solid line histogram), and the 583 galaxies with redshift based
solely on one or more emission lines (dotted-line histogram).}
\label{f-mag}
\end{figure}

This is illustrated in Fig.~\ref{f-mag} where we show the apparent
magnitude distribution of the 4447 galaxies with redshifts determined
only from absorption lines, of the 585 galaxies with redshifts
determined using both absorption and {\em emission lines}, and of the
583 galaxies for which the redshift is based only on emission lines
(for 19 of the 5634 galaxies magnitudes are not available). The
magnitude distribution of the galaxies with redshift based on {\em
emission lines only} is significantly different from the other two
(with $> .999$ probability, according to a Kolmogorov-Smirnov test,
see e.g. Press et al. 1986). This figure clearly illustrates the fact
that at faint magnitudes it is generally more difficult to obtain a
redshift from absorption lines than from emission lines.

From Fig.~\ref{f-mag} it is clear that the apparent fraction of ELG
varies considerably with magnitude. When calculating the intrinsic ELG
fraction one must take this magnitude bias into account (see also
\S~\ref{ss-bias}). However, the magnitude bias is unlikely to be
relevant in the analysis of the kinematics and the space distribution
of ELG and non-ELG. Since it has been established that velocities and
projected clustercentric distances are only very mildly correlated
with magnitude (see, e.g., Yepes, Dominguez-Tenreiro \& del Pozo-Sanz
1991, and Biviano et al. 1992, and references therein), it seems safe
to assume that the different magnitude distributions of ELG and
non-ELG will not affect our analysis of the observed space
distribution and kinematics.

The magnitude bias in Fig.~\ref{f-mag} could affect distributions of
clustercentric distance if in the ENACS the magnitude limit would vary
with distance from the cluster center. However, when we compare the
catalogues of cluster galaxies for which we obtained an ENACS redshift
with the (larger) catalogues of {\em all} galaxies brighter than our
magnitude limit (see Paper I), we find that no bias is present. In
other words: in all clusters that we observed in the ENACS the
completeness of the redshift determinations does not depend on
distance from the cluster center. This conclusion is strengthened by a
comparison of our spectroscopic catalogue with the nominally complete
photometric catalogues of Dressler (1980b), for the 10 clusters that
we have in common. Again, we detect no dependence of the completeness
on clustercentric distance.

We conclude therefore that the magnitude bias, which causes the
apparent fraction of ELG to increase strongly towards the magnitude
limit of the ENACS, only affects the estimation of the intrinsic ELG
fraction.  As is apparent from Fig.~\ref{f-mag}, that bias can be
avoided by restricting the analysis to the 585 ELG for which also an
absorption-line redshift could be obtained. However, it must be
realized that this remedy against the magnitude bias for ELG has one
disadvantage: it is likely to select against late-type spirals as
these occur preferentially in the class of ELG without absorption-line
redshift. We will come back to this in \S~\ref{ss-clusfld}.

For surveys of (cluster) galaxies at higher redshifts (and fainter
apparent magnitudes), which therefore have an inevitable observational
bias against galaxies without detectable emission lines, this bias can
in general {\em not} be corrected. Unless one has redshifts for all
galaxies, e.g. down to a given magnitude limit, conclusions drawn from
such `incomplete' samples can be seriously biased, as they refer
mostly to ELG. One obvious example is the determination of the
fraction of ELG as a function of redshift, but e.g. also the
determination of the evolution of cluster properties can be seriously
affected. This problem may be aggravated if, as we will discuss below
(see \S~\ref{s-spatial}), the spatial distributions of ELG and non-ELG
are not the same.

\section{The ELG fraction in clusters and the field}
\label{s-frac}

\subsection{Bias against Galaxies without Emission Lines}
\label{ss-bias}

\begin{figure}
\psfig{file=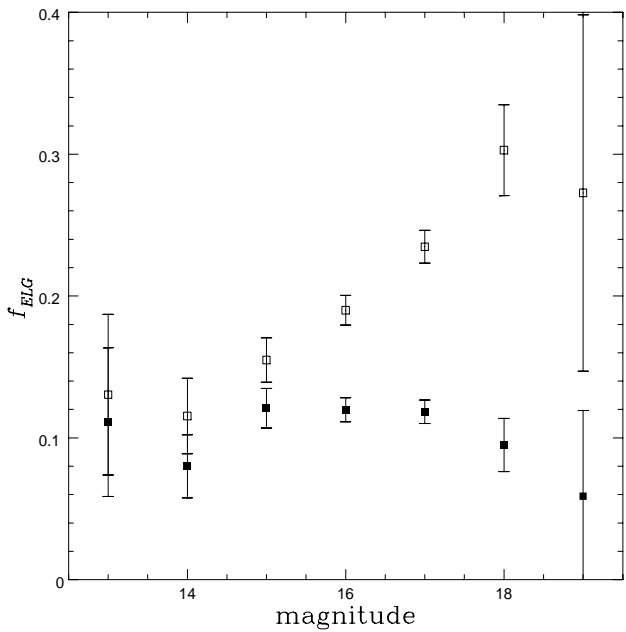,height=9.0cm,width=8.8cm}
\caption[]{The {\em apparent} fraction of ELG (open squares), 
determined using all galaxies, and the {\em true} fraction of ELG
(filled squares) determined as the fraction of galaxies that have
absorption- {\em and} emission-line redshifts among all galaxies with
absorption-line redshifts, as a function of magnitude (R$_{25}$).
Poissonian error bars are shown.}
\label{f-fracn}
\end{figure}

In Fig.~\ref{f-fracn} we show the fraction of ELG as a function of
apparent magnitude. The open symbols represent the {\em apparent} ELG
fraction, calculated as the total number of galaxies in the ENACS with
emission lines, divided by the total number of galaxies in the ENACS
in the same magnitude range, viz. as:

\begin{equation} 
f_{ELG} = \frac{\sum_{i=1}^{n} N_{ELG,i}}{\sum_{i=1}^{n} N_{i}}
\end{equation}

\noindent where $n$ is the number of systems, each containing $N_{i} 
\ge 10$ galaxies with redshifts, of which $N_{ELG,i}$ are ELG. The
strong increase of the apparent ELG fraction towards fainter
magnitudes is evident. As discussed in \S~\ref{ss-compl}, this
increase must be due to the bias that operates against the successful
determination of redshifts for faint galaxies without emission lines.

This bias can be overcome if we calculate the fraction of ELG as the
ratio of the number of the ELG for which also an absorption-line
redshift is available, by the total number of galaxies with an
absorption-line redshift. By definition, the magnitude bias does not
operate in this comparison. The filled symbols in Fig.~\ref{f-fracn}
give the resulting ELG fraction as a function of magnitude. As we
anticipated, there is essentially no dependence of this corrected,
{\em true} ELG fraction on magnitude, and it is considerably lower
than the apparent fraction, especially at fainter magnitudes. Only for
the brighter galaxies, for which there is no bias against the
detection of an absorption-line based redshift, are the apparent and
true ELG fraction essentially identical.

The apparent ELG fraction in the ENACS is 0.21 (=~1169~/~5634), but the
corrected value is 0.12 (=~586~/~5051). In this paper we will always
distinguish between the apparent and true ELG fractions, where the
latter is calculated from the sample of all galaxies with
absorption-line redshifts.

As far as we are aware, the correction for magnitude bias has not been
applied in earlier work on the ELG fraction.  In comparing our results
with other determinations this should always be kept in mind. It is
quite possible that some of the earlier results are not affected by
magnitude bias, but it is often difficult to find out if that is a
reasonable assumption. In a comparison with the results of the ESO
Slice Project (ESP, see e.g.\ Zucca et al.\ 1995), for which the same
instrumentation was used as for the ENACS, there may be differences in
bias which influence the result. The reason for this is that the
fraction of galaxies with emission lines is larger in the field (the
object of study in the ESP) than it is in our clusters.

All ELG fractions based on the ENACS include a small contribution from
AGN. Among interlopers and in systems with $N \le 3$ (which in the
ENACS provide the best approximation to the `field'), the AGN fraction
is $0.022\pm 0.006$. For the systems with $ N \geq 20$ (real, massive
clusters) it is only $0.007\pm 0.001$. These values are lower than the
values previously obtained by Dressler et al. (1985), Hill \& Oegerle
(1993), and Salzer et al. (1989, 1995), but this may be due (at least
partly) to the fact that we have been conservative in classifying
galaxies as AGN, and have probably accepted only those with the
strongest and broadest lines (see \S~\ref{ss-emiss}). The ratio of the
AGN fraction in the field and in clusters is $3 \pm 1$, consistent
with the value we find for all ELG (see \S~\ref{ss-clusfld}). Dressler
et al. (1985) found a similar value for the ratio between the AGN
fraction in the field and in clusters.

\subsection{The Fraction of ELG in Clusters and in the Field} 
\label{ss-clusfld}

The ELG fractions in clusters and field have been studied by several
authors, in order to find out if there is evidence for a difference in
the occurence of ELG which can be traced to the influence of the
environment in which galaxies live. Even though the ENACS, by its very
nature, does not contain many field galaxies, it contains a sufficient
number that we can investigate possible differences between the ELG
fractions in the field and in clusters.

It is not trivial to identify the field galaxies in the ENACS. The
main reason is that galaxies that are in small groups with only a few
measured redshifts could, on the one hand, be in the field but, on the
other hand, they could equally well be `tips of the iceberg'. In other
words: the number of {\em measured} redshifts in a group is not a good
criterion for assigning galaxies to the field or to a cluster. One
thing that is fairly certain is that the interlopers that were removed
from the systems on the basis of their position {\em and} velocity
(see \S~\ref{ss-systems}) belong to the field and we consider them to
be the best approximation to the field in the ENACS. Second best are
the isolated galaxies. Finally, galaxies in groups with at most 3
measured redshifts are acceptable candidates for field galaxies, since
the reality of such groups with less than 4 members is doubtful, as the
definition of systems with such a small number of members is not at
all robust (see Paper I). To a lesser extent the systems with 4 to
about 10 redshifts also do not have a very robust definition (ibid.)
but those we have included neither as cluster nor as field in the
comparison between field and clusters. Finally, systems with at least
10 measured redshifts are very likely to be real clusters or groups.

\begin{table}
\caption[]{The fraction of ELG in different environments}
\begin{flushleft}
\begin{tabular}{p{3.5cm} p{2.0cm} p{2.0cm}}
\hline
\noalign{\smallskip}
Environment & \multicolumn{2}{c}{f$_{ELG}$} \\
 & \multicolumn{1}{c}{apparent} & \multicolumn{1}{c}{true} \\
 & & \\
\hline
Interlopers               & $0.35 \pm 0.08$ & $0.22 \pm 0.07$ \\
Systems with $N \leq 3$   & $0.43 \pm 0.03$ & $0.21 \pm 0.02$ \\
Systems with $N \geq 10$  & $0.16 \pm 0.01$ & $0.10 \pm 0.01$ \\
\hline
\normalsize
\end{tabular}
\end{flushleft}
\label{t-data2}
\end{table}

In Tab.~\ref{t-data2} we show the resulting ELG fractions for the
three classes of environment. Note that for all three categories the
fractions have been calculated as in \S~\ref{ss-bias}. For each class
we have calculated the apparent as well as the true ELG fractions. The
ELG fractions for the interlopers and the $N \le 3$ systems are quite
similar, and they are both quite different from the average ELG
fraction in clusters. Because the galaxies for which the redshift is
based solely on emission lines have, on average, fainter magnitudes,
the difference appears most striking in the apparent fractions, but it
is equally significant in the bias-corrected, true values.

From the numbers in Tab.~\ref{t-data2} we conclude that it is not
unreasonable to assume that the interlopers and the galaxies in the $N
\le 3$ systems give a fair estimate of the ELG
fraction in the field: combining the two classes we obtain apparent
and true ELG fractions of $0.42 \pm 0.03$ and $0.21 \pm 0.02$
respectively. It is interesting to note that the corresponding numbers
for the systems with $4 \le N \le 9$ are $0.30 \pm 0.03$ and $0.15 \pm
0.02$. This clearly suggests that these systems are indeed
intermediate between real clusters and field galaxies. Additional
support for the assumption that the systems with $N \ge 10$ are indeed
almost all clusters is provided by the ELG fractions for the systems
with $N \ge 20$. For those, there is no doubt at all that they are
clusters and their average apparent and corrected ELG fractions are
$0.15 \pm 0.01$ and $0.10 \pm 0.01$ respectively.

Our {\em apparent} ELG fraction for the `field' is quite similar to
that derived by Zucca et al. (1995), who found a value of $\sim0.5$ in
the ESO Slice Project. This is quite gratifying, as these authors
obtained their spectra using an observational set-up that was
essentially identical to ours. It is true that the average redshift in
their survey is about a factor of 2 larger than in the ENACS, and
their result therefore applies to a larger region in the centre of the
galaxies than does ours. Apparently, this has little or no effect on
the apparent ELG fraction. The ELG fraction found by Salzer et al.
(1995) is 0.31, i.e. intermediate between our {\em apparent} and {\em
true} fractions. 

Our apparent ELG fraction for the field is significantly lower than
the value of $0.75 \pm 0.05$ that was found by Gisler (1978). On the
contrary, it is higher than the value of $0.31 \pm 0.05$ found by
Dressler et al. (1985), as well as the value of $0.27 \pm 0.08$ found
by Hill \& Oegerle (1993). However, as it is not clear whether we
should compare the literature values with our apparent or
bias-corrected values, the latter two determinations could actually be
consistent with our result.

A similar uncertainty is present in the comparison of our cluster ELG
fraction with earlier estimates in the literature. Our apparent value
of $0.16 \pm 0.01$ is consistent with the value found by Gisler (1978)
in compact clusters ($0.17 \pm 0.06$), but quite a bit higher than the
values of $0.07 \pm 0.01$ and $0.06 \pm 0.01$ found by Dressler et
al. (1985), and Hill \& Oegerle (1993), respectively. If the latter
two literature values should in fact be compared with our
bias-corrected value of $0.10 \pm 0.01$ the agreement becomes somewhat
better, although not perfect. As we shall see in \S~\ref{ss-veldisp},
part of the remaining difference in the cluster ELG fraction may be
due to the composition of the cluster samples with respect to mass (or
global velocity dispersion).

There are several other factors of this kind which can, at least in
principle, influence the observed ELG fraction. Among these are: the
average luminosity of the galaxy sample, the criterion by which
cluster members and field galaxies are identified, and (as mentioned
earlier) the linear sizes of the average aperture used in the
spectroscopy. The latter factor may well explain the differences with
the values obtained by Gisler (1978), who used spectra with a larger
effective aperture; this may be the reason for the systematically high
values that he obtained for the ELG fraction.  On the other hand, the
sample studied by Dressler et al. (1985) could be biased against
late-type spirals and irregulars (see Dressler \& Shectman 1988). As
these have a relatively high ELG fraction, this might well explain why
their ELG fractions (for cluster {\em as well as} for the field) are
low.

Although the {\em absolute} values of the ELG fractions obtained by
different authors may thus be difficult to compare (e.g. due to
differences in observational set-up etc.), the {\em relative}
fractions of ELG located in different environments might well be less
dependent on such details. In the ENACS the ratio between the ELG
fraction in the field and in clusters is $2.6 \pm 0.3$ (apparent) and
$2.1 \pm 0.3$ (bias-corrected). The average ratios found previously
are: $4.4 \pm 1.7$ (Gisler 1978), $4.4 \pm 1.0$ (Dressler et al. 1985)
and $4.5 \pm 1.4$ (Hill \& Oegerle 1993). The uncertainties are rather
large, but there may be some evidence that details of the various
techniques and the galaxy and/or cluster selection, have influenced
even the {\em relative} frequency of occurence of ELG in cluster and
field. On the other hand, the mix of the various types of galaxy may
not be the same in the different samples so that, with different ELG
fractions for the various galaxy types, the ratio between the
ELG fractions are expected to be different.

In Tab.~\ref{t-data3} we show the values of the ELG fraction for
galaxies in clusters as a function of morphological type. These
fractions are based on the ENACS data in combination with the
morphologies determined by Dressler (1980b) for the 545 galaxies in
the 10 clusters that are common between the ENACS and the Dressler
catalogue. Almost all of these (namely 537) have an absorption-line
ENACS redshift; 68 of the 537 galaxies (i.e. 13\%) also have emission
lines. Of the 68 ELG (none of which is an AGN), 60 are spirals or
irregulars, 7 are S0s and 1 is an elliptical. We thus find that the
fraction of ELG depends strongly on morphological type. Note that the
ELG fractions in Tab.~\ref{t-data3} are unbiased, as all galaxies
used in the statistics have absorption-line redshifts.

\begin{table}
\caption[]{The fraction of ELG for cluster galaxies of different 
           morphological types}
\begin{flushleft}
\begin{tabular}{ p{4.0cm} p{3.0cm}}
\hline
\noalign{\smallskip}
Morphological type & f$_{ELG}$ \\ & \\
\hline
E                   	& $0.01 \pm 0.01$ \\ 
S0                  	& $0.03 \pm 0.01$ \\ 
Sa, Sb              	& $0.27 \pm 0.05$ \\ 
Sc, Sd, I 		& $0.40 \pm 0.15$ \\ 
Unqualified S 		& $0.28 \pm 0.07$ \\
\hline
\normalsize
\end{tabular}
\end{flushleft}
\label{t-data3}
\end{table}

It is also of interest to determine the fraction of spirals that we
have detected as ELG. Of the 180 spirals in the sample of 537
galaxies, only 60 are ELG. So, while most of our ELG are late-type
galaxies, the ELG represent only about $1/3$ of the total spiral
population in our clusters.

Since the mix of galaxy types is a strong function of the density of
the environment, one may ask whether the difference between the ELG
fractions in the clusters and in the field can be totally attributed
to a lower fraction of late-type galaxies in clusters. Following
Dressler et al. (1985), we have used the ELG fractions of cluster
galaxies for the different galaxy types, and convolved that with the
distribution over galaxy type of field galaxies. This should yield the
ELG fraction that clusters would have if their morphological mix were
the same as that of the field. In the ENACS there are only very few
field galaxies with known type. Therefore, we have assumed the field
type mix given by Oemler (1974), with which we calculate an expected
field ELG fraction of $0.23 \pm 0.03$. This value, which is based on
the assumption that the dependence of ELG fraction over morphological
type is identical in cluster and field, is of course fully consistent
with our observed, bias-corrected value for the field ELG fraction.

This result is at variance with all previous findings on this point
(Osterbrock 1960, Gisler 1978, Dressler et al. 1985, Hill \& Oegerle
1993). It can be rephrased by saying that environmental effects
probably do not affect the fraction of ELG, or the emission-line
activity. Note that, had we not accounted for the magnitude bias (the
fact that the apparent ELG fraction increases towards faint
magnitudes), we would have come to the same conclusion as the
above-mentioned authors. However, the magnitude bias is stronger for
the field galaxies than for the cluster sample (because our field
galaxies are on average fainter than our cluster galaxies). As a
result, the need for different emission-line characteristics of field
and cluster galaxies disappears if the bias is taken into account.

\begin{figure}
\psfig{file=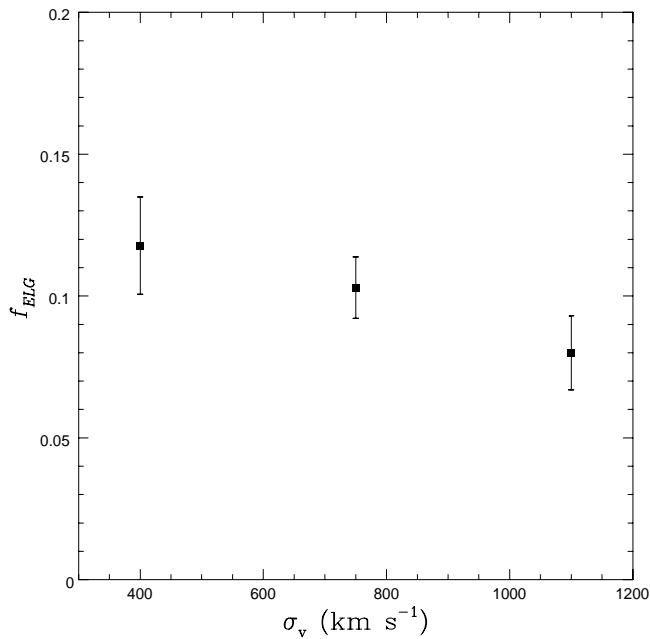,height=9.0cm,width=8.8cm}
\caption[]{The fraction of ELG in systems with different velocity 
dispersions, \sv.  Poissonian error bars are shown.}
\label{f-fracsig}
\end{figure}

At this point we must come back to the selection against late-type
spirals which is inherent in our calculation of the true ELG fraction,
since the latter is based only on the ELG {\em with} absorption-line
redshift (see \S~\ref{ss-compl}). We have attempted to take this factor
into account, by assuming that most of the ELG {\em without}
absorption-line redshift in the field are late-type spirals. Our best
estimate of the fraction of late-type spirals among our field spirals
is about 50\%, although we cannot exclude that it is 70\%. Using the
former fraction together with the ELG fractions for early- and
late-type spirals in Tab.~\ref{t-data3}, we estimate an expected ELG
fraction in the field of $0.26 \pm 0.05$ instead of $0.23 \pm 0.03$.
This is still consistent with the observed true ELG fraction in the
field of $0.21 \pm 0.02$, so that the conclusion in the preceding
paragraph is not likely to be the result of the selection against
late-type spirals in the calculation of true ELG fractions.

\subsection{ The ELG fraction as a function of velocity dispersion}
\label{ss-veldisp}

In \S~\ref{ss-clusfld} we found that $f_{ELG}$ is practically independent
of $N$ for systems with $N \geq 10$. On the other hand, we also noted
that some of the differences between our ELG fractions and those of
other authors might be due to different composition of cluster samples
in terms of mass, or some other physically relevant parameter. An
obvious question is therefore if, within the ENACS data, we observe a
dependence of the ELG fraction on the global velocity dispersion of
the system. In Fig.~\ref{f-fracsig} we show $f_{ELG}$ as a function of
velocity dispersion, where $f_{ELG}$ was calculated as in
\S~\ref{ss-bias} in three separate intervals of $\sv$. For this
figure, we have used only the 75 systems with $N \ge 20$ of sample 3,
as these are very likely to be bona-fide rich clusters. It is clear
that there is a significant decrease of the ELG fraction with
increasing velocity dispersion, by a factor of 1.5 over the range of
dispersions sampled. Within the errors, the same result is obtained if
we use the sample of all 120 systems with $N \ge 10$ listed in
Tab.~\ref{t-data1}.

{\em On average}, clusters with smaller velocity dispersions are less
rich than clusters with larger velocity dispersions (see, e.g., Paper
II). Since essentially all ELG are spirals, the above result is thus
consistent with van~den~Bergh's (1962) finding that the fraction of
late-type galaxies is higher in poorer clusters. We must point out
that the $f_{ELG}$ dependence on \sv is not induced by different sizes
of the area over which we obtained spectroscopy for the different
clusters. This could have an effect, in principle, as a consequence of
the morphology-density relation and because the clusters for which the
observations covered a larger area have a (slightly) higher \sv than
average. However, if we consider only galaxies within 1~\Mpc of their
respective cluster center, in those 51 clusters (with $N \ge 20$)
observed at least out to 1~\Mpc, the relation between $f_{ELG}$ and
\sv is unchanged.

We conclude therefore that there is a significant decrease of the
fraction of ELG, with increasing \sv, which must reflect a dependence
on mass.

\section{The global kinematics of ELG and non-ELG}
\label{s-kine}

In this section we analyze the global kinematics of ELG and
non-ELG. Before we enter into the details of this discussion we want
to emphasize the following important point. All the results that we
will obtain in this section, either on differences between ELG and
non-ELG in average velocity or in velocity dispersion, are based on
the implicit assumption that both types of galaxies consist of single
systems. In other words: we have calculated a single average velocity
(or velocity dispersion) for both ELG and non-ELG. If this assumption
is incorrect (e.g. because the ELG do not have a smooth spatial
distribution, but instead are in several compact groups within a
cluster) the interpretation of the results obviously becomes more
complicated. We will return to this question in \S~\ref{s-substr}.

\begin{figure}
\psfig{file=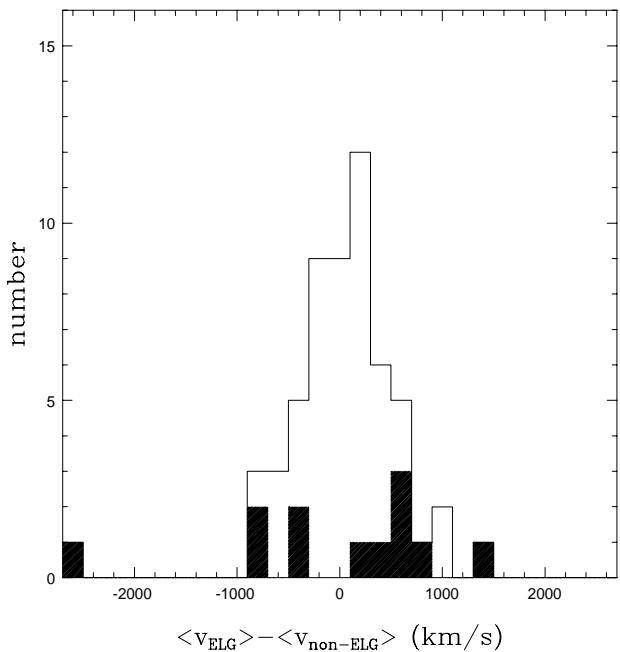,height=9.0cm,width=8.8cm}
\caption[]{The distribution of \vme$-$\vma
for the 57 clusters with at least 5 ELG within $\pm 3 \sigma$ from the
mean velocity. The twelve clusters for which this difference is
significant at more than $2 \sigma$ have been indicated.}
\label{f-vdiff}
\end{figure}

\subsection{Average velocities}
\label{ss-average}

\begin{figure}
\psfig{file=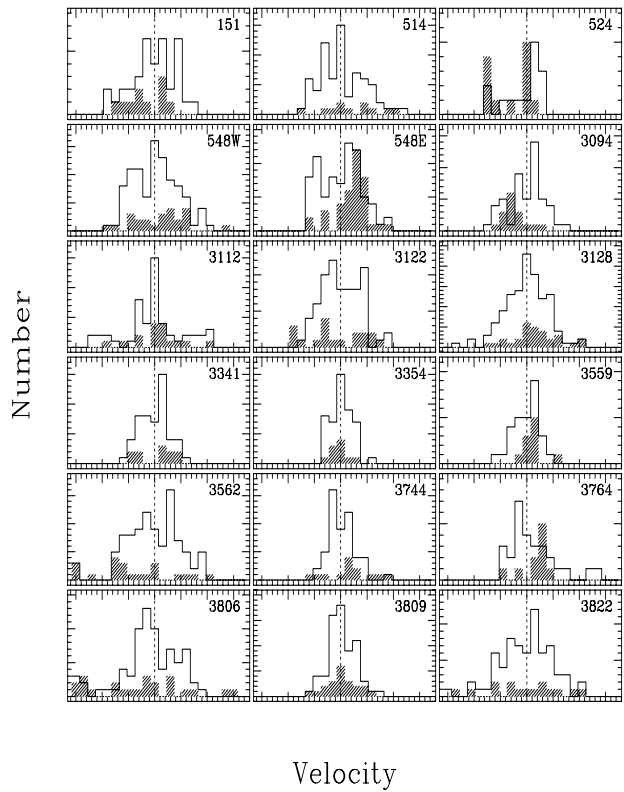,height=11.0cm,width=8.8cm}
\caption[]{Velocity Distributions of non-ELG and ELG (hatched histogram)
in the 18 clusters with at least 10 ELG. The dashed line in each panel
indicates the average velocity of the system. One division on the
horizontal (velocity-) scale corresponds to 200 \ks and the binwidth
is 250 \ks. One division on the vertical scale corresponds to one
galaxy.}
\label{f-3cls}
\end{figure}

\begin{table}
\caption[]{The average velocity differences 
between ELG and non-ELG in those clusters 
where the difference is larger than $2 \sigma$}
\begin{flushleft}
\small
\begin{tabular}{r p{1.3cm} p{0.3cm} p{0.5cm} p{1.1cm} p{0.3cm} p{0.5cm} rr}
\hline
\noalign{\smallskip}
Abell & \multicolumn{3}{c}{\vma} & \multicolumn{3}{c}{$\Delta V$} 
      & \multicolumn{2}{c}{$N,N_{ELG}$} \\
\multicolumn{1}{c}{nr} & \multicolumn{3}{c}{\ks} 
                       & \multicolumn{3}{c}{\ks} & & \\
\noalign{\smallskip}
\hline
\noalign{\smallskip}
 151\ \ \ & \multicolumn{1}{r}{12122} & $\pm$ & \multicolumn{1}{r}{112} 
          & \multicolumn{1}{r}{-340}  & $\pm$ & \multicolumn{1}{r}{161}
          & 25 & 5 \\
 151\ \ \ & \multicolumn{1}{r}{29537} & $\pm$ & \multicolumn{1}{r}{165}
          & \multicolumn{1}{r}{367}   & $\pm$ & \multicolumn{1}{r}{175}
          & 35 & 5 \\
 548E     & \multicolumn{1}{r}{12268} & $\pm$ & \multicolumn{1}{r}{99} 
          &  \multicolumn{1}{r}{530}  & $\pm$ & \multicolumn{1}{r}{153}
          & 114 & 38 \\
 548\ \ \ & \multicolumn{1}{r}{25186} & $\pm$ & \multicolumn{1}{r}{459}
          & \multicolumn{1}{r}{1496}  & $\pm$ & \multicolumn{1}{r}{588}
          & 14 & 8 \\
 548\ \ \ & \multicolumn{1}{r}{30081} & $\pm$ & \multicolumn{1}{r}{122} 
          & \multicolumn{1}{r}{ 575}  & $\pm$ & \multicolumn{1}{r}{220} 
          & 21 & 6 \\
2819\ \ \ & \multicolumn{1}{r}{22239} & $\pm$ & \multicolumn{1}{r}{74}  
          & \multicolumn{1}{r}{  243} & $\pm$ &  \multicolumn{1}{r}{106}
          & 49 & 6 \\
2819\ \ \ & \multicolumn{1}{r}{25889} & $\pm$ & \multicolumn{1}{r}{52} 
          & \multicolumn{1}{r}{ -712} & $\pm$ &  \multicolumn{1}{r}{313}
          & 43 & 6 \\
3094\ \ \ & \multicolumn{1}{r}{20155} & $\pm$ & \multicolumn{1}{r}{124} 
          & \multicolumn{1}{r}{ -489} & $\pm$ &  \multicolumn{1}{r}{163} 
          & 66 & 16 \\
3151\ \ \ & \multicolumn{1}{r}{20414} & $\pm$ & \multicolumn{1}{r}{143} 
          & \multicolumn{1}{r}{-2555} & $\pm$ &  \multicolumn{1}{r}{583}
          & 38 & 6 \\
3562\ \ \ & \multicolumn{1}{r}{14744} & $\pm$ & \multicolumn{1}{r}{123} 
          & \multicolumn{1}{r}{ -862} & $\pm$ &  \multicolumn{1}{r}{382}
          & 116 & 21 \\
3693\ \ \ & \multicolumn{1}{r}{26887} & $\pm$ & \multicolumn{1}{r}{213} 
          & \multicolumn{1}{r}{  791} & $\pm$ &  \multicolumn{1}{r}{268}
          & 16 & 5 \\
3764\ \ \ & \multicolumn{1}{r}{22329} & $\pm$ & \multicolumn{1}{r}{206} 
          & \multicolumn{1}{r}{  555} & $\pm$ &  \multicolumn{1}{r}{231}
          & 38 & 10 \\
\hline
\normalsize
\end{tabular}
\end{flushleft}
\label{t-goodvdif}
\end{table}

Zabludoff \& Franx (1993) noted that the average velocity of late-type
galaxies was different from that of early-type galaxies in 3 of the 6
clusters they examined. They interpreted this as evidence for
anisotropic infall of groups of spirals into the cluster. However,
since their analysis is limited to 6 clusters, one cannot draw general
conclusions from this result.

Here, we address the same issue on the basis of our sample of 57
clusters in which at least 5 ELG were found (sample~1). For these
systems, we determined the average velocities of both ELG and non-ELG,
as well as the associated 1$\sigma$ errors, which were calculated with
the jack-knife technique (see, e.g., Beers et al. 1990). For the 12
clusters in which the velocity difference between ELG and non-ELG
exceeds $2 \sigma$, we give details in Tab.~\ref{t-goodvdif}. The
distribution of the differences \vme$-$\vma in the 57 systems is shown
in Fig.~\ref{f-vdiff}.

We thus find a much lower fraction of clusters with significant
differences in the average velocities of ELG and non-ELG than did
Zabludoff \& Franx. One might think that the two results could be
consistent, if in many of our clusters there would be a real
difference which has been masked by the effects of limited statistics.
However, in \S~\ref{ss-vdisp} we will show, from the distributions of
velocity difference between galaxy pairs, that this is unlikely to be
the case. In addition, the same low fraction of significant velocity
offsets is found among the 20 systems that contain at least 10 ELG.

For each of the 18 systems with at least 10 ELG (sample 2), we show in
Fig.~\ref{f-3cls} the velocity distributions of ELG and non-ELG
separately. Note that this figure does not include every system listed
in Tab.~\ref{t-goodvdif}, because quite a few of those have less than
10 ELG. For the 4 systems in the figure that also appear in
Tab.~\ref{t-goodvdif} (A548E, A3094, A3562 and A3764) the histograms
clearly give a visual confirmation of the existence of a velocity
difference. There are several systems with intrigueingly uneven
velocity distributions for, in particular the ELG, but with the
present statistics it is impossible to say if those are indeed
clusters with real velocity differences between ELG and non-ELG.

\subsection{Velocity dispersions}
\label{ss-vdisp}

For the systems with significant velocity differences between ELG and
non-ELG that are shown in Fig.~\ref{f-3cls}, the numerical evidence is
supported visually by the figure. However, it is impossible to say
from that figure if there exist significant differences between the
{\em velocity dispersions} of ELG and non-ELG. It turns out, however,
that among the 18 systems with at least 10 ELG, 3 have a \sv
difference between ELG and non-ELG that is significant at a level of
more than 2 $\sigma$. The values of \sva and (\sve - \sva) for these
systems and their jack-knife errors are given in cols.(2) and (3) of
Tab.~\ref{t-goodsigv}. It is interesting that all 3 differences are
positive, i.e. that in all 3 cases the \sv of the ELG is larger than
that of the non-ELG.

We have followed up this conclusion by considering all 75 systems of
sample 3 with $N \ge 20$. For 57 of these, it is not possible to
derive a meaningful \sv estimate for the ELG separately. However, for
71 of the 75 systems (4 of which do not have an ELG), one can compare
the \sv values derived for the total galaxy population with those for
the non-ELG only (i.e. excluding the ELG). As non-ELG are the dominant
population, we expect that the change in \sv on excluding the ELG will
be quite small, but combining the results for all 71 systems may
nevertheless give a significant result.

\begin{table}
\caption[]{Significant velocity-dispersion differences between ELG
           and non-ELG}
\begin{flushleft}
\small
\begin{tabular}{r p{1.3cm} p{0.3cm} p{0.5cm} p{1.1cm} p{0.3cm} p{0.5cm} rr}
\hline
\noalign{\smallskip}
Abell & \multicolumn{3}{c}{\sva} & \multicolumn{3}{c}{$\Delta \sv$} 
      & \multicolumn{2}{c}{$N,N_{ELG}$} \\
\multicolumn{1}{c}{nr} & \multicolumn{3}{c}{\ks} 
                       & \multicolumn{3}{c}{\ks} & & \\
\noalign{\smallskip}
\hline
\noalign{\smallskip}
3122 & \multicolumn{1}{r}{706} & $\pm$ & \multicolumn{1}{r}{59} 
     & \multicolumn{1}{r}{354} & $\pm$ & \multicolumn{1}{r}{119} 
     & 89 & 18 \\
3744 & \multicolumn{1}{r}{474} & $\pm$ & \multicolumn{1}{r}{55} 
     & \multicolumn{1}{r}{519} & $\pm$ & \multicolumn{1}{r}{80} 
     & 66 & 13 \\
3806 & \multicolumn{1}{r}{953} & $\pm$ & \multicolumn{1}{r}{113}
     & \multicolumn{1}{r}{763} & $\pm$ & \multicolumn{1}{r}{275} 
     & 97 & 23 \\
\hline
\normalsize
\end{tabular}
\end{flushleft}
\label{t-goodsigv}
\end{table}

\begin{figure}
\psfig{file=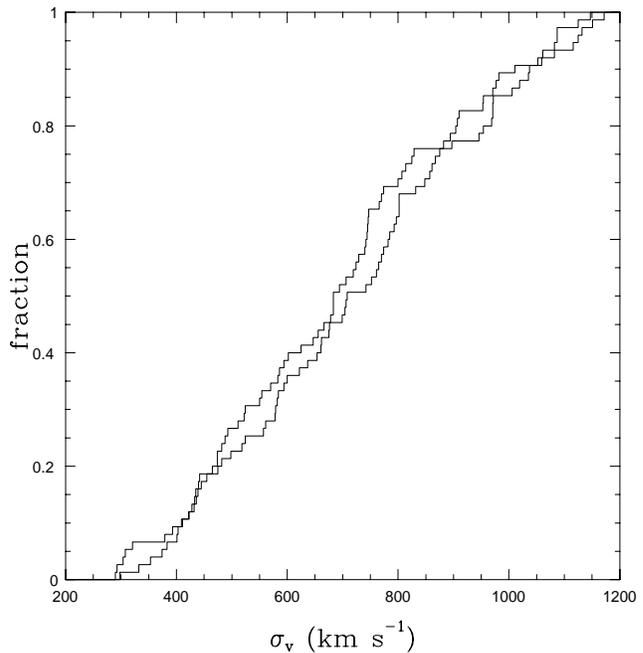,height=9.0cm,width=8.8cm}
\caption[]{The cumulative \sv
distributions for non-ELG only (thin line), and for all the galaxies
(ELG$+$non-ELG, thick line) in the 75 clusters with at least 20
members.}
\label{f-scum}
\end{figure}

In Fig.~\ref{f-scum} we show the two cumulative \sv-distributions for
all the galaxies (ELG$+$non-ELG) and for non-ELG only, in the 75
clusters with at least 20 members (i.e. the 4 clusters without ELG are
included in the Figure). The removal of the ELG from the cluster
samples in general lowers the value of \sv; a Wilcoxon test (see e.g.
Press et al.  1986) indicates that the ELG$+$non-ELG \sv distribution
is different from that of the non-ELG at the $> 0.999$~conf.level, and
that \sv of ELG$+$non-ELG is, on average, larger than \sv of the
non-ELG.

\subsection{Velocity distributions}
\label{ss-vdist}

In order to examine the kinematical properties of ELG and non-ELG
further, we have put together all galaxies in the 75 clusters in
sample 3 in a single, `synthetic', cluster. We define a ``normalized
velocity difference'', \vn, with respect to the average velocity of
the system to which the galaxy belongs, which is normalized with
respect to the velocity dispersion of the parent system, viz.  
$ \vn = (v- <v>) / \sv $.

For this discussion we could have included the systems with $10
\leq N < 20$, but we have not done so, because later on we will
include positional information which requires the centre to be known
with sufficient accuracy. When comparing the \vn-distributions of ELG
and non-ELG, we do not want to be strongly affected by the tails of
these distributions. As our interloper rejection method was only
applied to clusters with more than 50 galaxies (see
\S~\ref{ss-systems}), it is possible that a few outliers are still
present in the systems with less than 50 galaxies. For the preceding
analysis, in which we used robust estimators, such outliers were not
very important. However, combining data for many systems for which the
average velocity is not known exactly will produce longer tails in the
velocity distribution. As for some of the following analyses we cannot
use robust estimators we have to get rid of possible outliers. To that
end we have applied a 3$\sigma$-clipping criterion (Yahil \& Vidal
1977). This removes 30 galaxies in total (among which are 9 ELG) and
yields a `synthetic' cluster with 3699 galaxies, among which are 549
ELG.

\begin{figure}
\psfig{file=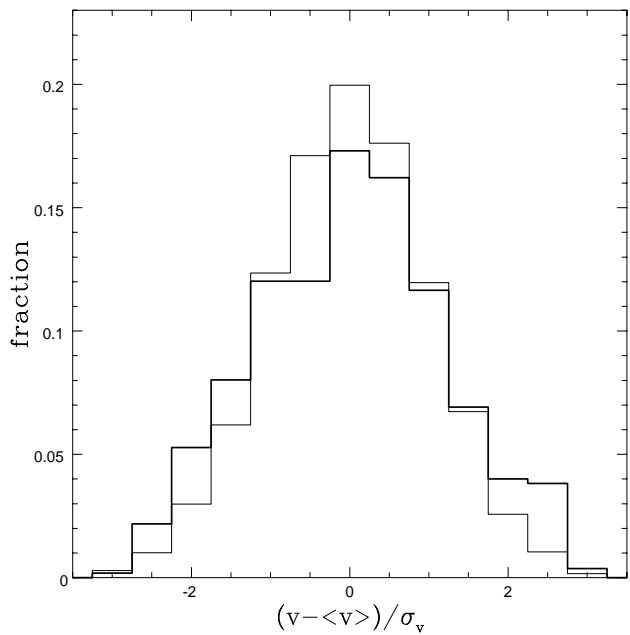,height=9.0cm,width=8.8cm}
\caption[]{The normalized-velocity histograms for the total sample
of 549 ELG (thick line) and 3150 non-ELG (thin line) in the 75
clusters with at least 20 members.}
\label{f-vnorm}
\end{figure}

The ELG and non-ELG \vn-distributions are shown in Fig.~\ref{f-vnorm}.
The \vn-distribution for ELG is broader than that for non-ELG; the
KS-test gives a probability of 0.029 that the two distributions are
drawn from the same parent population. The dispersion of the \vn's of
the ELG is $21 \pm 2$~\% larger than the dispersion of the \vn's of
the non-ELG.

Among the 549 ELG, 37 are AGN; the KS-test indicates that the
\vn-distributions of AGN and non-ELG are significantly different (with a
probability of 0.047 for the two distributions to be drawn from the
same parent population), but the \vn-distributions of AGN and the
other 512 ELG are not. Therefore, AGN seem to follow the velocity
distribution of the other ELG.

In principle there are two possible explanations for the wider \vn
distribution for the ELG. On the one hand, the ratio of \sve and \sva
may be {\em larger than unity} by roughly the same amount in {\em
essentially all} systems. On the other hand, the broader distribution
of the \vn of the ELG could be due to the fact that we have superposed
many ELG systems. Even if, in most systems, the \sv of the ELG were
identical to the \sv of the non-ELG, the width of the \vn distribution
could be larger for ELG than for non-ELG if the {\em average
velocities} of ELG and non-ELG are {\em substantially different} in
the large majority of the systems. The reason is that the \vn's are
calculated with the overall values of $<\mbox{v}>$ and $\sigma_V$,
which are determined primarily by the non-ELG.

\begin{figure}
\psfig{file=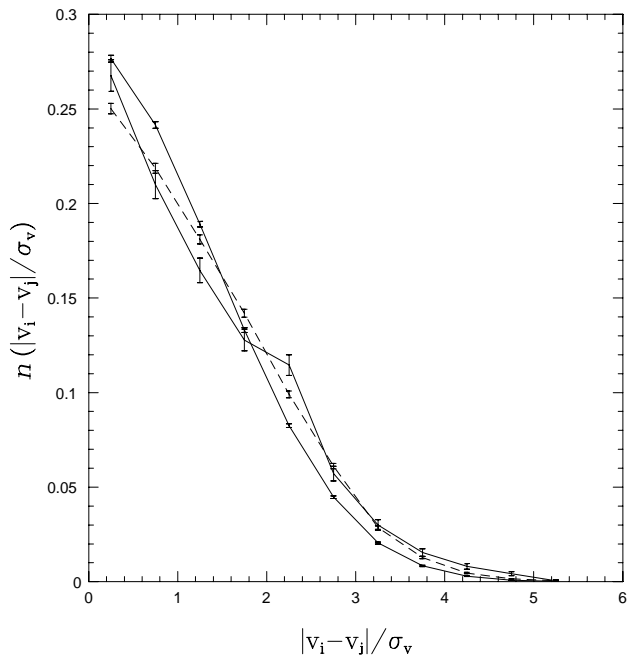,height=9.0cm,width=8.8cm}
\caption[]{The distribution of velocity differences among pairs
of non-ELG (thin line), pairs of ELG (thick line), and mixed pairs
of one non-ELG and one ELG (dashed line), normalized to the
velocity dispersion of the system to which the pair belongs.
Poissonian error-bars are shown.}
\label{f-ndeltav}
\end{figure}

These two possible explanations are obviously extreme cases, and it is
very unlikely that one of them applies exclusively. In
\S~\ref{ss-average} we saw that in a small fraction of the  clusters 
there is evidence for a significant offset between the average
velocities of ELG and non-ELG. However, we could not tell whether such
offsets occur in essentially all systems (but were not detectable in
many systems due to limited statistics). Here, we will show that the
main reason for the apparently larger \sv of the ELG must be that the
intrinsic \sv of the ELG is about 20\% larger than that of the
non-ELG. In other words: only a small part of the larger dispersion of
the ELG is due to the fact that we have combined several narrower
gaussians with different means. This conclusion is based on an
analysis of the pairwise velocity differences of the ELG.

In Fig.~\ref{f-ndeltav} we show the sum (over all systems) of the
distribution, for all galaxy pairs in a given system, of the absolute
value of the pairwise velocity difference (again, normalized to the
velocity dispersion of the system to which the two galaxies belong),
i.e. $\mid v_i-v_j \mid /\sv$. These distributions were calculated
separately for pairs of non-ELG (thin line), pairs of ELG (thick
line), and for mixed pairs of an ELG and a non-ELG (dashed line). The
three distributions have been normalized to the total number of pairs
of each kind; clearly the uncertainties are largest for the
ELG/ELG-pairs. If essentially all ratios \sve / \sva for the
individual systems would be larger than one (and if velocity offsets
between ELG and non-ELG were non-existent) one would expect three
gaussian distributions in Fig.~\ref{f-ndeltav}, with widths increasing
from the non-ELG/non-ELG, via the non-ELG/ELG to the ELG/ELG pairs.

The distributions for the non-ELG/non-ELG and non-ELG/ELG pairs are
indeed very close to gaussian, and the non-ELG/ELG distribution is
broader than that of the non-ELG/non-ELG pairs. However, the
distribution for the ELG/ELG pairs is quite different from this
gaussian expectation. The ELG/ELG distribution for $\Delta$v smaller
than $\approx 2$ has a curvature opposite to that of a gaussian.
Therefore, there must be a component that produces an n($\mid v_i-v_j
\mid /\sv$) that is small for small values of $\mid v_i-v_j \mid /\sv$
and has a peak at $\mid v_i-v_j \mid /\sv$ of about 2 and then
decreases again. One way to produce such a component is by having
systems in which the ELG have a velocity offset of about one \sv
w.r.t. to the non-ELG. However, at the same time, there must be a
second component which produces the broadening of the ELG/ELG
distribution for large values of $\mid v_i-v_j \mid /\sv$ (say, larger
than about 2). In other words: we are led to a schematic model with
two components in the ELG velocity distribution, one with fairly small
internal \sv and significant velocity offsets, and another with a
global \sv that is larger than the \sv of the non-ELG but without a
significant velocity offset.

We have attempted to estimate the relative importance of these two
components by some simple modeling. Although there is not a single,
unique solution, it appears that the distribution for the ELG/ELG
pairs in Fig.~\ref{f-ndeltav} requires that $\sim$25\% of the ELG
reside in systems with an average velocity offset of about 600 \ks
(i.e. almost equal to the value of the global $\sigma_V$). However,
the internal \sv of these ELG systems with significant velocity
offsets must be small, i.e. less than about half the value of
\sva. If the fraction of ELG in these systems is much larger or
smaller than 25\% and/or the \sv values of these systems is comparable
to the \sv values of the non-ELG, the steep slope of the ELG/ELG
distribution at small $\mid v_i-v_j \mid /\sv$ values (say, below 1.5)
cannot be reproduced.

For the other $\sim$75\% of ELG, i.e. those in the systems without
large velocity offsets, the global value of \sv must be a factor of
about 1.25 larger than \sva in order to reproduce the number of
ELG/ELG pairs for values of $\mid v_i-v_j \mid /\sv$ between 2 and 4
to 5. This simple model clearly cannot give information on how the
latter 75\% of ELG are distributed, and how their \sve (of, on
average, 1.25 \sva) comes about. As mentioned earlier, they can either
be essentially isolated galaxies (and distributed more or less
uniformly in their parent clusters), or they may be in compact groups,
or a combination of these. From Fig.~\ref{f-3cls} one gets the
impression that both cases occur. We will return to this question in
\S~\ref{s-substr}. 

It is worth remembering that in \S~\ref{ss-average} we found that for
12 out of 57 systems there is a significant difference in the average
velocities of ELG and non-ELG. The observed offsets range from about
300 to 1400 \ks (with a median of about 600 \ks). Both the fraction of
systems with a significant offset and the size of the offsets that we
derived here from a simple model thus agree very nicely with the
observed values.

Finally, we note that the distribution of normalized velocities for
the AGN subset of the ELG cannot be distinguished from that of the
non-ELG or ELG, due to the limited number of AGN in the ENACS.

\section{The spatial distributions of ELG and non-ELG}
\label{s-spatial}

We have analyzed the spatial distributions of ELG and non-ELG (and
possible differences between them) in several different ways. First,
we have used the harmonic mean pair distances, \rh, for which no
cluster centre needs to be known. As the number of ELG per system is
often not very large, the determination of \rh for the ELG separately
is mostly not very robust. We have therefore compared the cumulative
\rh distributions of all cluster galaxies (ELG$+$non-ELG) and of 
non-ELG only, for the 75 systems of sample 3. According to the
Wilcoxon test, the two distributions are significantly different (at
the $> .999$~conf.level). More specifically: when ELG are excluded
from the systems, smaller \rh values are found. Although these
differences are systematic, they are quite small because the average
fraction of ELG is only 16 \%. The average reduction of \rh is only
3~\% which implies that \rhe is larger than \rha by $\sim 20$~\%.

\begin{figure}
\psfig{file=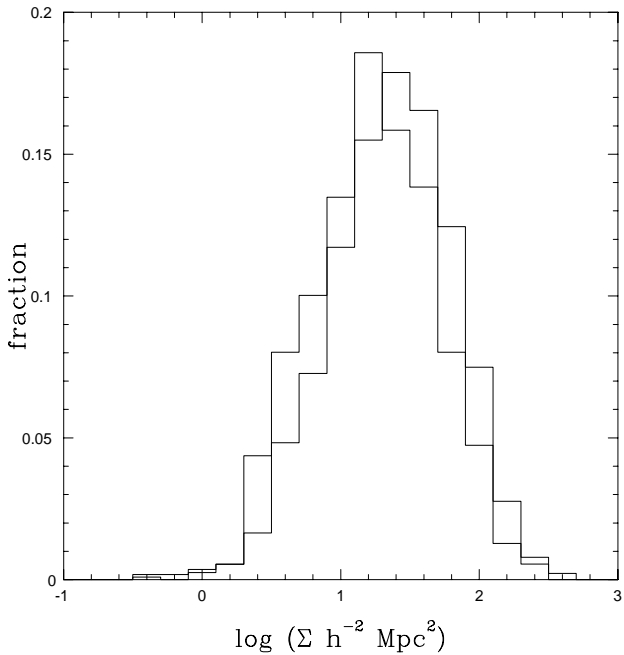,height=9.0cm,width=8.8cm}
\caption[]{The distribution of the logarithm of the local surface
densities, $\Sigma$, for ELG (thick line) and non-ELG (thin line) in 75
clusters with $N \geq 20$. See the text for the definition of local
densities.}
\label{f-rho}
\end{figure}

Another way to look at the differences in the spatial distribution of
ELG and non-ELG is to study the local densities of their immediate
environment. We calculated the local density, $\Sigma$, as the surface
density of galaxies within a circular area centered on the galaxy,
with radius equal to the distance to its N$^{1/2}$-th neighbour, where
$N$ is the total number of galaxies in the system. In Fig.~\ref{f-rho}
we show the normalized distributions of the values of $\Sigma$ for ELG
and non-ELG, for all galaxies in the 75 systems of sample 3. The
distributions are significantly different (at the $>
0.999$~conf.level), and the local density around ELG is, on average,
$0.72 \pm 0.03$ times the density around non-ELG.

Both tests show that the spatial distribution of the ELG is
significantly broader than that of the non-ELG. Additional information
on the differences between the spatial distributions of ELG and
non-ELG can be obtained from a comparison of the density profiles of
the two classes.  Because the number of ELG in a cluster is rather
small, a reliable density profile of the ELG can only be obtained from
the combination of all systems. We then assume implicitly that
different clusters have similar profiles. This is not unreasonable,
since cluster density profiles have similar slopes (see, e.g., Lubin
\& Bahcall 1993, Girardi et al. 1995), although their core-radii have
a large spread (see, e.g., Sarazin 1986; Girardi et al. 1995; note,
however, that the very existence of cluster cores is doubtful, see,
e.g., Beers \& Tonry 1986, Merritt \& Gebhardt 1995). All these
possible complications are not very important at this point, because
here we are only interested in the relation between the density
profiles of ELG and non-ELG.

In constructing the surface density profiles we have considered only
the 51 systems from sample 3 for which the data extend at least out
to 1~\Mpc, in order to avoid possible problems of incompleteness, and
we have limited our analysis to galaxies within 1~\Mpc.

\begin{figure}
\psfig{file=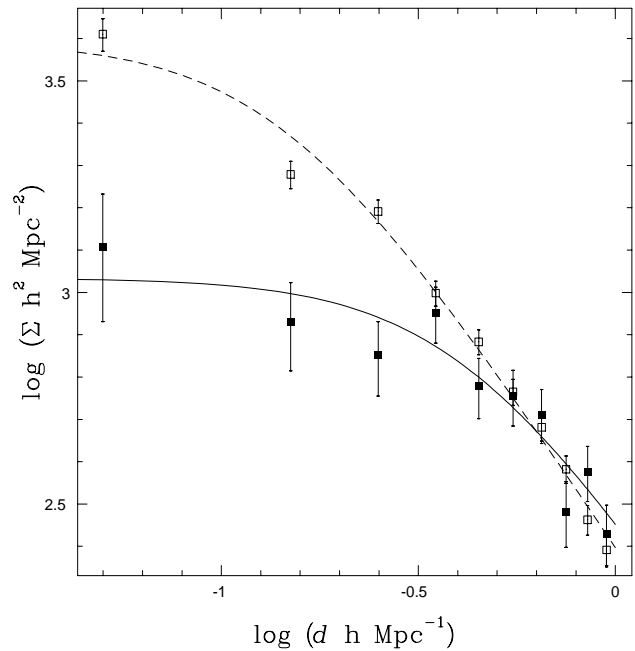,height=9.0cm,width=8.8cm}
\caption[]{The surface density profiles for ELG (filled symbols) and 
non-ELG (open symbols) for 51 clusters sampled at least out to 1~\Mpc,
and with at least 20 galaxy members. The continuous and dashed lines
are the fits to the ELG and non-ELG distributions, respectively with
$\beta=-0.71$. Note that the ELG profile has been moved up by $+0.65$
in $\log \Sigma$ for an easier comparison with the non-ELG profile.}
\label{f-dprof}
\end{figure}

The density profiles of ELG and non-ELG are shown in Fig.~\ref{f-dprof},
and they have been fitted by the usual $\beta$-model:
\begin{equation}
\Sigma(d)=\Sigma(0)[1+(d/r_{c})^2]^{-\beta}
\end{equation}
The maximum-likelihood fit to the unbinned distribution of the non-ELG
yields the following values: $\beta=-0.71 \pm 0.05$, r$_c = 0.15 \pm
0.04$~\Mpc, with a reduced $\chi^2$ of 1.9 (8 degrees of freedom). For
the ELG we obtain maximum-likelihood values $\beta=-1.3 \pm 1.2$,
r$_c = 0.8 \pm 0.8$~\Mpc, with a reduced $\chi^2$ of 0.9 (again, 8
degrees of freedom). The simultaneously fitted model-parameters for
the ELG are quite uncertain, largely due to the flatness of the ELG
density profile within 1~\Mpc. We have therefore made a second fit to
the ELG data in which we have taken $\beta=-0.71$ (equal to the value
for the non-ELG), which gives r$_c = 0.42 \pm 0.07$~\Mpc for the ELG.

The $\beta$-models with $\beta=-0.71$ are also shown in
Fig.~\ref{f-dprof}. The fit for the non-ELG is not very good because
of the peak in the first bin (note that a peaky profile is expected
when an accurate choice of the cluster center is made; see Beers \&
Tonry 1986).  Nevertheless, the values found for r$_c$ and $\beta$ are
consistent with recent results obtained by Lubin \& Bahcall (1993)
and Girardi et al. (1995). 

We note in passing that the AGN, which are a subset of the ELG, have a
spatial distribution that cannot be distinguished from that of the ELG;
however, their distribution is different from that of the non-ELG.

\section{Correlations between velocity and position}
\label{s-substr}

In \S~\ref{s-kine} and \S~\ref{s-spatial} we discussed separately the
kinematics and spatial distribution of ELG and non-ELG and the
differences between them. From the discussion in \S~\ref{ss-vdist} we
concluded that there is evidence for two ELG populations, one with a
\sv that is considerably smaller than the overall value and with
significant velocity offsets (w.r.t. the non-ELG), and another with
\sv larger than the overall value and without significant velocity
offsets. This result immediately raises the question of possible
correlations between velocity and position or, in other words: of
structure in phase-space. Do the characters of the phase-space
distributions of ELG and non-ELG differ and if so, in what way. What
evidence do we have on substructure, i.e. on the existence of
spatially and/or kinematically compact groups, and are there
differences between ELG and non-ELG in that respect.

\subsection{The phase-space distributions}
\label{ss-phase}

\begin{figure}
\psfig{file=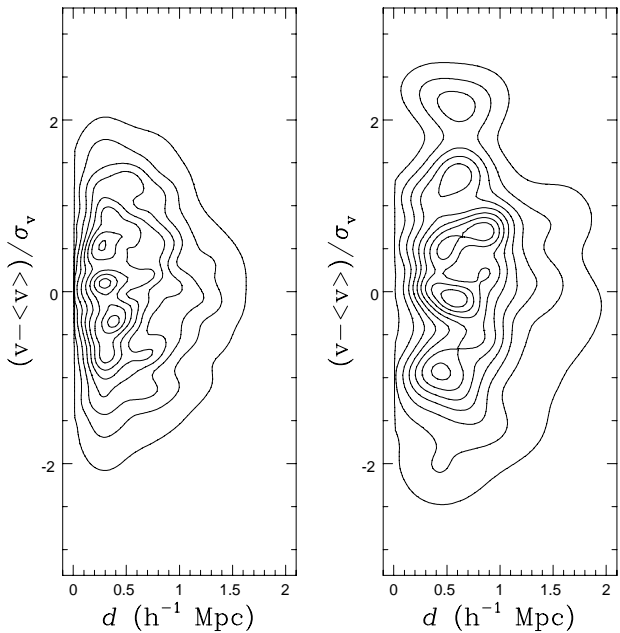,height=9.0cm,width=8.8cm}
\caption[]{Adaptive-kernel maps of the 2-dimensional distribution
w.r.t normalized velocity and clustercentric distance for the non-ELG
(left panel) and ELG (right panel) in the synthetic cluster
constructed from the 75 systems with $N \geq 20$.}
\label{f-disvel}
\end{figure}

In Fig.~\ref{f-disvel} we show adaptive kernel maps (see e.g. Merritt
\& Gebhardt 1995) of the distributions of both ELG and non-ELG w.r.t. 
normalized-velocity (see \S~\ref{ss-vdist}) and clustercentric
distance, for the synthetic cluster constructed from the 75 systems
with $N \geq 20$. Note that a velocity limit of $\pm 3$\sv has been
applied, as before. A 2~-~D KS-test (Fasano \& Franceschini 1987)
gives a probability $< 0.001$ that the two distributions are drawn
from the same parent distribution. This is hardly surprising in view
of the fact that we found a less centrally concentrated spatial
distribution for the ELG than for the non-ELG, as well as a \sv that
is $\sim$ 20\% larger for the (majority of the) ELG than it is for the
non-ELG. Both effects are clearly visible in Fig.~\ref{f-disvel}.
However, it is very difficult to tell which features in the
distributions in Fig.~\ref{f-disvel} represent real substructure, if
only because the distributions represent sums over all 75 clusters.
It is equally difficult to estimate from Fig.~\ref{f-disvel} what
fraction of the galaxies is in real substructure that is compact both
in position and velocity.

For a more quantitative discussion of this point we consider the
distributions of $\Delta$r$_{proj}$ and \vn for {\em pairs} of
galaxies (rather than individual galaxies) and, in particular, pairs
of {\em {\bf nearest} neighbours from the same class}. For the non-ELG
we use all 75 systems in sample 3 (with $N \ge 20$) which contain 3150
galaxies in total. The number of non-ELG nearest-neighbour pairs is
2219. This is less than the number of galaxies because when B is the
nearest neighbour of A {\em and}, at the same time, A happens to be
the nearest neighbour of B, the pair A-B is used only once. For the
ELG we have considered only the 18 systems with N$_{ELG}\ge$10 (for
reasons that will become apparent); these 18 systems contain 306 ELG
(3 ELG were removed in the $\pm 3\sigma$ clipping) with which
we have formed 207 nearest-neighbour pairs.

\begin{figure}
\psfig{file=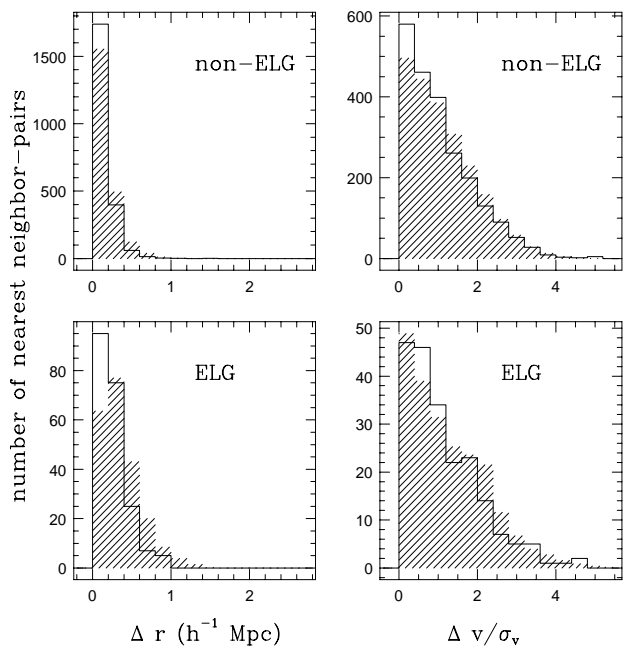,height=9.0cm,width=8.8cm}
\caption[]{The observed distributions of $\Delta{r_{proj}}$ and 
$\Delta v/\sv$ for all nearest-neighbor pairs of non-ELG (top) and ELG
(bottom) are shown as full-drawn line histograms. The shaded
histograms represent the same distributions obtained from the
observations after scrambling w.r.t. radial velocity and azimuthal
angle (see text for details).}
\label{f-dvdr}
\end{figure}

In Fig.~\ref{f-dvdr} we show the normalized distributions of
$\Delta$r$_{proj}$ and \vn (i.e. $\Delta v/\sv$) for nearest
neighbours, for non-ELG (upper two panels) and ELG (lower two
panels). The global differences between the two sets of distributions
are not unexpected: the lower surface density of ELG gives rise to
larger $\Delta$r$_{proj}$ for ELG-ELG pairs; similarly, the larger
global \sv of the ELG causes a wider \vn distribution for the ELG-ELG
pairs. In order to get a more quantitative estimate of the amount of
real, compact substructure in Fig.~\ref{f-disvel}, we have compared
these distributions with scrambled versions of the same. The scrambled
data should give the number of accidental pairs with given values of
$\Delta$r$_{proj}$ and \vn, and thus show what fraction of the
structure in Fig.~\ref{f-disvel} is real. The shaded histograms in
Fig.~\ref{f-dvdr} represent the $\Delta$r$_{proj}$ and \vn
distributions for scrambled versions of the ELG and non-ELG datasets.

In principle, the scrambling of the (r,v)-datasets can be done in
three ways. First, one may leave the values of r$_{proj}$ and v
intact, and only reassign the value of the azimuthal angle of each
galaxy randomly. This will keep both the radial density profile as
well as the \sv -profile intact. However, in that case the galaxies
near the centre of a system (with small values of r$_{proj}$, and
consequently also small values of $\Delta$r$_{proj}$) globally retain
their relative velocities, and the scrambling will be far from
perfect. Secondly, one may apply velocity scrambling. In that case,
the \sv-profile is not conserved; however, the average decrease of \sv
over 1 \Mpc is modest (see, e.g. den Hartog and Katgert 1996), and we
do not consider the non-conservation of the \sv-profile a serious
problem.

However, if one does not scramble the azimuthal angle at the same
time, velocity scrambling only makes sense if the number of galaxies
in a system is quite large. If that is not the case, there will be an
important amount of `memory' between the pairs in the original and in
the scrambled data. Therefore, we applied both velocity- and azimuth
scrambling. Even then, the scrambled ELG distribution may have
significant memory of the observed distribution in view of the small
average number of ELG (and therefore ELG-ELG nearest-neighbour pairs)
in a system. To minimize this effect (which will lead to an
underestimation of the amount of real small-scale structure) we have
used for the ELG only the 20 systems with at least 10 ELG (remember
that for the non-ELG we used the 75 systems with at least 20 members).

From Fig.~\ref{f-dvdr} we conclude that both for the non-ELG and the
ELG there is an excess of nearest-neigbour pairs with
$\Delta$r$_{proj} <$ 0.2 h$^{-1}$ Mpc, viz. of about 7\% for the
non-ELG and about 15\% for the ELG. Moreover, for the non-ELG there
appears to be a small excess (of about 4\%) of nearest-neighbour pairs
with \vn $\la$ 0.6. For the ELG the excess is about 7 \% , but the
values of \vn are between $\approx$ 0.5 and 1.2. The number of excess
pairs in the \vn distribution is about half that in the
$\Delta$r$_{proj}$ distribution, for ELG as well as non-ELG. This must
mean that there is more `memory' about velocity than about position in
the scrambled datasets. Nevertheless, it seems safe to conclude from
Fig.~\ref{f-dvdr} that the ELG show more small-scale structure than
the non-ELG. However, whereas the non-ELG excess pairs have small
$\Delta$r$_{proj}$ as well as small \vn, the ELG excess pairs have
small $\Delta$r$_{proj}$ but fairly large \vn's.

We are thus led to a picture in which a fairly small fraction of the
galaxies are in 'real' pairs with small $\Delta$r$_{proj}$ and \vn,
with the fraction of ELG in such pairs probably slightly larger
($\approx 20\%$) than that of non-ELG ($\approx 10\%$). Interestingly,
the estimated fraction of ELG in pairs is quite consistent with the
value derived in \S~\ref{ss-vdist}. It is a bit puzzling that we now
find that the \vn's of these pairs are not very small, whereas in
\S~\ref{ss-vdist} we found that \sv for these ELG must be quite small.
If one assumes these ELG pairs to be in groups, and if one assumes the
relation between the average $\Delta$v and \sv, valid for a gaussian,
to hold for those putative groups, one derives typical masses of
several times $10^{12}$ solar masses (using the projected virial mass
estimator for isotropic orbits, see Heisler, Tremaine \& Bahcall
1985). This implies that the real ELG pairs could be in small groups
of a few to several ELG, depending on the average mass of the ELG in
question.

\subsection{Substructure}
\label{ss-dressler}

It is interesting to find out whether the groups of ELG (and, to a
lesser extent, non-ELG) that we `detected' in the analysis in
\S~\ref{ss-phase}, are detectable as substructure in the 
velocity-position databases of individual clusters as well. In order
to investigate this we have applied the test (due to Dressler \&
Shectman 1988, but with the modifications proposed by Bird 1994) for
the presence of substructure. This test compares the value of a
substructure parameter, $\Delta = \sum_{i=1}^{N} \delta_{i}$, for a
cluster, with the distribution of values of the same parameter that
one obtains in 1000 Monte Carlo randomizations of the cluster
data-set. A large value of $\delta_i$ for a given galaxy implies a
high probability for it to be located in a spatially compact
subsystem, which has either a \vm that differs from the overall
cluster mean, or a different \sv, or both.

We have applied this test to the 25 systems with $N \geq 50$. These
contain a sufficiently large number of galaxies (on average 86 of
which 14 are ELG) that for these systems the test may be expected to
produce significant results. An additional advantage of this selection
is that from all these systems interlopers were removed. In
Tab.~\ref{t-subst} we list the probability P$_{\Delta}$ that a value
of $\Delta$ as large as the one observed is obtained by chance. When
this probability is low, one thus has strong evidence for
subclustering. The probability P$_{\Delta}$ was calculated separately
for all galaxies (ELG+non-ELG) (col.3), and for the non-ELG only
(col.4), i.e. with the ELG removed.

\begin{table}
\caption[]{The Dressler \& Shectman test for substructure}
\begin{flushleft}
\small
\begin{tabular}{rcrrrr}
\hline
\noalign{\smallskip}
name & $<z>$ & \multicolumn{2}{c}{P$_{\Delta}$} & 
               \multicolumn{2}{c}{N,N$_{ELG}$} \\
     &       & \multicolumn{1}{c}{all} & \multicolumn{1}{c}{non-ELG} & & \\
\noalign{\smallskip}
\hline
\noalign{\smallskip}
  119 \ \ \  &0.044& 0.620 & 0.742 & 101 & 5  \\
  168 \ \ \  &0.045& 0.324 & 0.277 &  76 & 6  \\
  514 \ \ \  &0.072& 0.017 & 0.048 &  81 &11  \\
 548W        &0.042& 0.000 & 0.003 & 120 &24  \\
 548E \      &0.041& 0.000 & 0.003 & 114 &38  \\
  978 \ \ \  &0.054& 0.129 & 0.071 &  61 & 7  \\
 2734 \ \ \  &0.062& 0.063 & 0.095 &  77 & 1  \\
 3094 \ \ \  &0.068& 0.000 & 0.000 &  66 &16  \\
 3112 \ \ \  &0.075& 0.241 & 0.688 &  67 &16  \\
 3122 \ \ \  &0.064& 0.005 & 0.021 &  89 &18  \\
 3128 \ \ \  &0.060& 0.000 & 0.000 & 152 &30  \\
 3158 \ \ \  &0.059& 0.491 & 0.218 & 105 & 9  \\
 3223 \ \ \  &0.060& 0.179 & 0.042 &  73 & 6  \\
 3341 \ \ \  &0.038& 0.579 & 0.546 &  63 &11  \\
 3354 \ \ \  &0.059& 0.004 & 0.000 &  57 &10  \\
 3558 \ \ \  &0.048& 0.247 & 0.235 &  73 & 9  \\
 3562 \ \ \  &0.048& 0.000 & 0.019 & 116 &21  \\
 3651 \ \ \  &0.060& 0.021 & 0.061 &  78 & 8  \\
 3667 \ \ \  &0.056& 0.212 & 0.306 & 103 & 9  \\
 3695 \ \ \  &0.089& 0.001 & 0.000 &  81 & 9  \\
 3744 \ \ \  &0.038& 0.061 & 0.025 &  66 &13  \\
 3806 \ \ \  &0.076& 0.078 & 0.201 &  97 &23  \\
 3809 \ \ \  &0.062& 0.072 & 0.032 &  89 &21  \\
 3822 \ \ \  &0.076& 0.064 & 0.053 &  84 &15  \\
 3825 \ \ \  &0.075& 0.072 & 0.114 &  59 & 4  \\
\hline\normalsize
\end{tabular}
\end{flushleft}
\label{t-subst}
\end{table}

In 8 systems we find evidence for substructure at the 0.99~conf.level,
using all galaxies (i.e. for A548W, A548E, A3094, A3122, A3128, A3354,
A3562 and A3695). In addition, there are 2 systems with substructure
at the 0.98~conf.level, viz. A514 and A3651. One might suspect that
the systems with a substructure signal are preferentially found among
the systems with the largest number of galaxies, as for those it will
be relatively easier to detect deviations from uniformity. From
Table~\ref{t-subst} it indeed appears that there is small effect of
this kind: the 8 systems with P$_{\Delta}$ less than 0.01 have an
average number of galaxies of $98 \pm 13$, whereas for the other 17
systems this number is $79 \pm 4$.

Perhaps more significantly, the 8 systems with signs of substructure
have $20 \pm 4$ ELG, and the other 17 systems only $9 \pm 1$ ELG on
average. This might lead one to suspect that the ELG are a very
important, if not {\it the}, cause of substructure.  However, there is
no evidence that that is so; among the 8 systems with P$_{\Delta} <
0.01$ for all galaxies, there are 6 for which P$_{\Delta}$ is still
less than 0.01 if the ELG are excluded.  Therefore, it is very
unlikely that the presence of ELG is a requirement for the occurrence
of substructure.

However, there is an indication that the ELG preferentially occur in
substructure, if the system to which the ELG belong indeed does have
substructure. This presumes that the substructure is probably
delineated primarily by the non-ELG (and/or the dark matter), and that
the ELG so to speak `follow' the substructure that is present. This
conclusion is based on the following evidence.

Combining all systems with $N \geq 50$, we have compared the
distributions of the individual values of $\delta_i$ of the 1808
non-ELG and 340 ELG in these 25 systems. According to a KS-test, the
probability that the distributions are drawn from the same population
is $< 0.001$. Note that this conclusion does not depend critically on
the 8 clusters with clear evidence of substructure. When we exclude
these clusters, the $\delta_i$ distributions of ELG and non-ELG are
still different at the 0.994~conf.level. Using the total sample of
galaxies in the 25 clusters, we find that the fraction of ELG is
almost twice as large among the galaxies that, according to their
value of $\delta_i$, are more likely to reside in substructure, than
among the galaxies that are not likely to belong to substructure;
$f_{ELG} = 0.15 \pm 0.02$ for galaxies with $\delta_i \geq 2$, and
$f_{ELG} = 0.08 \pm 0.02$, for galaxies with $\delta_i
\leq 1$.

We conclude therefore that in substructures the ELG occur {\em
relatively} more frequently than the non-ELG. In a fairly small
fraction of the systems they may even account for most of the
substructure; however, in general the ELG seem to {\em follow} the
substructure rather than that they {\em define} it. As we saw in
\S~\ref{ss-clusfld} there is a clear tendency for the fraction of
ELG to be larger in smaller systems. It is thus not totally unexpected
to find that the ELG are relatively more associated with substructure
since, to some extent, the ELG can be regarded as low-richness groups
within richer systems. While ELG are more frequently found in
subclusters than non-ELG, their average velocities are seldom
different from the cluster ones; therefore, groups containing ELG
cannot be rapidly infalling into the cluster, unless the infall of
these groups is more or less isotropic.

\section{Non-equilibrium and orbits of ELG}
\label{s-formev}

For the 75 systems with $N \geq 20$ we have computed the virial and
projected masses (see, e.g. Heisler et al. 1985); we have done this
for the datasets that include all members as well as for the subsets
of non-ELG members. For both mass estimators we find that the estimate
based on all the galaxies is 8~\% larger than that based on non-ELG
only. The distribution of masses computed using only non-ELG is
significantly different (at the $>$0.999~conf.level) from that
computed from the combination of non-ELG and ELG.

We have estimated the average ratio of the masses we would have
derived separately for ELG and non-ELG. Using an average ELG fraction
of 0.15 for the 75 systems used here, we estimate that cluster mass
estimates based solely on ELG must, on average, be $\sim$50~\% larger
than the estimates based on the non-ELG only. Note that this result
involves the assumption that ELG and non-ELG have the same type of
orbital distribution, so that the same velocity projection factor
applies. If the orbital characteristics of ELG and non-ELG are not the
same, the difference in the mass estimates may in reality be larger or
smaller.

In deriving the difference of 50~\% in estimated mass, it has also
been tacitly assumed that ELG and non-ELG are both pure categories.
However, one must realize (see also \S~\ref{ss-clusfld}) that the
non-ELG class may harbour a significant contribution of late-type
galaxies (about two-thirds of all spirals were not detected as
ELG). If the latter share the kinematics of the ELG, there might be an
even larger difference between mass estimates based on spiral and
non-spiral galaxies.

To forge consistency between the mass estimates based on ELG and
non-ELG, the orbits of the non-ELG should be more radial than those of
the ELG in order to counteract the lower value of ${\sv}^2$ by a
larger velocity projection factor. However, this is very unlikely in
view of the more centrally concentrated distribution of the non-ELG.
Another, more probable solution to the apparent inconsistency between
the mass estimates based on ELG and non-ELG is to assume that the ELG
are not in equilibrium with the non-ELG. As both classes are in the
same potential (to which both probably contribute in a limited way),
it would only seem possible for them not to be in equilibrium if they
had not adjusted to the potential in the same manner. This could
happen if their relaxation times were very different, which in turn
could be a natural consequence of the differences in their spatial
distribution. As the non-ELG are significantly more concentrated and
find themselves in a denser environment, they are more likely to have
reached equilibrium than are the ELG.

We have tried to obtain more information on the orbits of ELG and
non-ELG by analyzing the dependence on projected radius of the
distribution of the line-of-sight component of their velocities. If
the statistics of the orbital parameters of ELG and non-ELG are
different, their velocity distributions must depend on position in
different ways, and that would manifest itself in the distribution of
line-of-sight velocities (see e.g. Kent \& Gunn 1982, and Merrit
1987). 

We have determined the radial dependence of the dispersion of the
line-of-sight velocity, using the synthetic cluster formed by adding
the 75 systems with at least 20 members. This has the clear advantage
of statistical weight but the equally clear disadvantage of producing
an `average' cluster that may bear little resemblance to any of the
real clusters that it is made up of. After clipping of the $> 3
\sigma$ outliers this cluster contains 3699 galaxies of which 549 are
ELG. In Table~\ref{t-sigmas1} we show the values of \sv for two radial
bins as well as the overall value, for ELG and non-ELG. The bins have
been chosen as a compromise between optimizing the detection
probability for orbital anisotropy, if it exists, and the statistical
weight for its detection. E.g., decreasing the size of the inner bin
increases the discrimating power for orbital anisotropy, but decreases
the statistical weight for its detection.
 
\begin{table}
\caption[]{The observed radial dependence of $\sigma_V$}
\begin{flushleft}
\small
\begin{tabular}{crrcc}
\hline
\noalign{\smallskip}
r$_{proj}$ & \multicolumn{2}{c}{number of gals} &
	     \multicolumn{2}{c}{\sv$_{,obs}$} \\
\noalign{\medskip}
Mpc & non-ELG & ELG & non-ELG & ELG \\
\hline
\noalign{\medskip}
$<$ 0.75 & 2170 & 310 & 1.05 & 1.28 \\
$>$ 0.75 &  980 & 239 & 0.92 & 1.13 \\
all      & 3150 & 549 & 1.01 & 1.22 \\
\noalign{\smallskip}
\hline
\end{tabular}
\end{flushleft}
\label{t-sigmas1}
\end{table}

In adding the data for the 75 systems, the values of r$_{proj}$ have
not been scaled but the velocities have been scaled with the global
value of \sv of each cluster. The values of \sv in
Tab.~\ref{t-sigmas1} are thus in units of the overall \sv of the
synthetic cluster. One might wonder why \sv of the non-ELG is larger,
instead of smaller, than 1.00 (after all, the combination of ELG and
non-ELG should give \sv exactly equal to 1.00). The reason is that, in
adding many normalized guassians, the errors in the means of the
individual guassians produce an overall sigma that is slightly larger
than 1.00. 

The Table confirms the large difference between ELG and non-ELG as
regards overall \sv and, at the same time, shows that the ratios of
the \sv's in the inner and outer bin are remarkably similar (viz. 1.14
and 1.13) for ELG and non-ELG. The first impression could be that this
indicates similar orbits for ELG and non-ELG.  However, that cannot be
the case, as the density distributions of ELG and non-ELG are
significantly different.

From some fairly simple modeling, we have predicted values of \sv for
ELG and non-ELG in the two radial bins defined in
Tab.~\ref{t-sigmas1}. We used a model with 3~-~D density profiles
described by $\beta$-models with a $\beta$ of -0.71, and r$_{\rm c}$'s
of 0.15 and 0.42 \Mpc for non-ELG and ELG respectively, i.e the values
we found from the fits to the surface density profiles (see
\S~\ref{s-spatial}). In addition, we specify a velocity distribution
at each position that is described by a value for the velocity
dispersion in the radial direction, $\sigma_{rad}$(r), and a constant
anisotropy parameter ${\cal A}$ (= 1--$(\sigma_{tan}/\sigma_{rad})^2$).
The value of $\sigma_{rad}$(r) was assumed to depend linearly on
radius, viz. $\sigma_{rad}$(r) = $\sigma_{rad}$(0) - $\varpi$ r.

From the 3-D density profile, we randomly extract 10$^5$ points. Since
our clusters are sampled to different limiting radii, the total sample
including all clusters is not complete at large distances from the
center. To mimic this incompleteness in our simulated sample, we did
not include the simulated points with r$_{proj} >$ 1.2~\Mpc. From
Tab.~\ref{t-data1} is can be seen that the median of the largest
r$_{proj}$ in the clusters is essentially 1.2~\Mpc. We included all
points with r$_{proj} <$ 1.2~\Mpc as long as the 3~-~D clustercentric
distance was less than 10~\Mpc. In this manner, we reproduced to
within a few percent the observed fractions of galaxies in the inner
and outer bins, for ELG as well as non-ELG.

For each of the simulated points, we randomly extracted three velocity
components from three gaussian velocity distributions that follow from
the velocity dispersion profile and the anisotropy. We assumed the two
components in the tangential directions to have the same dispersion,
which follows from $\sigma_{rad}$ and ${\cal A}$. We then projected
the velocity vector along the line-of-sight, and added noise to the
resulting line-of-sight velocity by adding two random deviates. The
first simulates the errors in the individual velocity measurements
(assumed to be 0.1 \sv), the other the error in the average cluster
velocity (assumed to be 0.2 \sv). Finally, as in the observations, we
rejected all simulated (line-of-sight) velocities outside $\pm 3
\sigma$.

\begin{table*}
\caption[]{The modeled radial dependence of \sv}
\begin{flushleft}
\small
\begin{tabular}{|c|c|c|c|c|cccccc|}
\hline
 & \multicolumn{2}{c|}{non-ELG} & \multicolumn{8}{c|}{ELG} \\
\cline{2-11}
              & obs & model & obs     & \multicolumn{7}{c|}{model} \\
\cline{2-11}
$\cal A$          & & 0     & & 0     & -0.9  & -0.6  & -0.3  & 0.3   
                                      & 0.6   & 0.9   \\
$\sigma_{rad}$(0) & & 1.13  & & 1.50  & 1.25  & 1.31  & 1.38  & 1.64  
                                      & 1.86  & 2.34  \\
$\varpi$          & & -0.14 & & -0.19 & -0.16 & -0.16 & -0.17 & -0.21 
                                      & -0.23 & -0.29 \\
\hline
$<$ 0.75 & 1.05 & 1.05 & 1.28 & 1.26 & 1.25 & 1.25 & 1.25 & 1.26 & 1.27 & 
   1.28 \\
$>$ 0.75 & 0.92 & 0.92 & 1.13 & 1.17 & 1.19 & 1.18 & 1.17 & 1.17 & 1.15 & 
   1.13 \\
  all    & 1.01 & 1.01 & 1.22 & 1.22 & 1.22 & 1.22 & 1.22 & 1.22 & 1.22 & 
   1.22 \\
\hline
\end{tabular}
\end{flushleft}
\label{t-sigmas2}
\end{table*}

The model parameter were optimized as follows. First, we estimated the
best values of $\sigma_{rad}$(0) and $\varpi$ for the non-ELG,
assuming the orbits to be isotropic, i.e. $\cal A \equiv$~0.0. In
order to reproduce the observed values of \sv for the non-ELG we need
$\sigma_{rad}$(0) = 1.13 and $\varpi$ = --0.14. This implies that
$\sigma_{rad}$ decreases to zero at a radial distance of 8.1 Mpc,
which is quite acceptable in view of the expected turn-around radius
of the `synthetic' cluster. This model produces inner and outer \sv's
for the non-ELG of 1.05 and 0.92 respectively, i.e. exactly as
observed.

Next, we tried to model the observed \sv values of the ELG, again
assuming isotropic orbits. We have no prescribed relation between the
$\sigma_{rad}$(0)'s of ELG and non-ELG, except that it is very hard,
if not impossible, to imagine that the ratio of the
$\sigma_{rad}$(0)'s for ELG and non-ELG could exceed $\sqrt 2$. The
best value of $\sigma_{rad}$(0) for the ELG was found to be 1.50,
which then implies a value $\varpi$ = --0.19 (because we assume that
the radius at which $\sigma_{rad}$ decreases to zero is the same for
ELG and non-ELG). This model does not do a very bad job, but it does
not fully reproduce the decrease of \sv from the inner to the outer
bin. One way to improve the agreement between model and observations
would be to assume a larger value for $\varpi$ for the ELG than for
the non-ELG. Although we cannot totally exclude the possibility that
the ELG have a steeper velocity dispersion profile than the non-ELG,
the data that we have for the inner part of the `synthetic' cluster
do not indicate this (see also below).

Another way to improve the agreement between observed and predicted
\sv's of the ELG is to assume that the velocity distribution of the
ELG is anisotropic; in other words: to assume that the anisotropy
parameter $\cal A \neq$~0. In that case there are two free parameters:
$\sigma_{rad}$(0) and $\cal A$; the value of $\varpi$ will follow
directly from $\sigma_{rad}$(0) and the maximum radius of 8.1 Mpc
found for the non-ELG. The best solution has to be determined by
iteration because the observed global value of \sv does not, for a
given value of $\cal A$ (assumed to be independent of radial
distance), immediately yield the value of $\sigma_{rad}$(0). If
$\sigma_{1d}$ had been derived in an aperture with sufficiently large
projected radius, {\em and} if $\varpi$ = 0, $\sigma_{rad}$(0) would
have followed directly as $\sqrt{3/(3-2 \times {\cal A})} \times
\sigma_{1d}$. As neither of these two conditions is fulfilled, we had 
to iteratively find the best combination of $\sigma_{rad}$(0) and
$\varpi$ for each value of $\cal A$ that we assumed.

In Tab.~\ref{t-sigmas2} we summarize the results of the modeling, for
non-ELG as well as ELG. It can be seen that models with
$(\sigma_{tan}/\sigma_{rad}) >$ 1.0 (i.e $\cal A <$ 0) predict that
the observed dispersion of the line-of-sight velocity component does
not decrease sufficiently with r$_{proj}$. This was to be expected
since we needed the anisotropy to increase the ratio between the inner
and outer values of \sv of the ELG, but that can only be accomplished
with radially elongated orbits.

The most probable value of $\cal A$ for the ELG is not very well
determined from our data, because the uncertainties in the observed
inner and outer values of \sv for the ELG are estimated to be 4--5\%.
However, taking the data at face-value we conclude that the best fit
is obtained for $\cal A \approx$ 0.9, but a value of 0.3, or even 0.0
cannot be formally exluded. On the contrary, we think that negative
values of $\cal A$ can be fairly safely ruled out. From
Tab.~\ref{t-sigmas2} we therefore conclude that $\cal A \approx$
0.5$\pm$0.5, which thus provides some evidence for anisotropy of the
ELG orbits but not very strong evidence.

\begin{figure}
\psfig{file=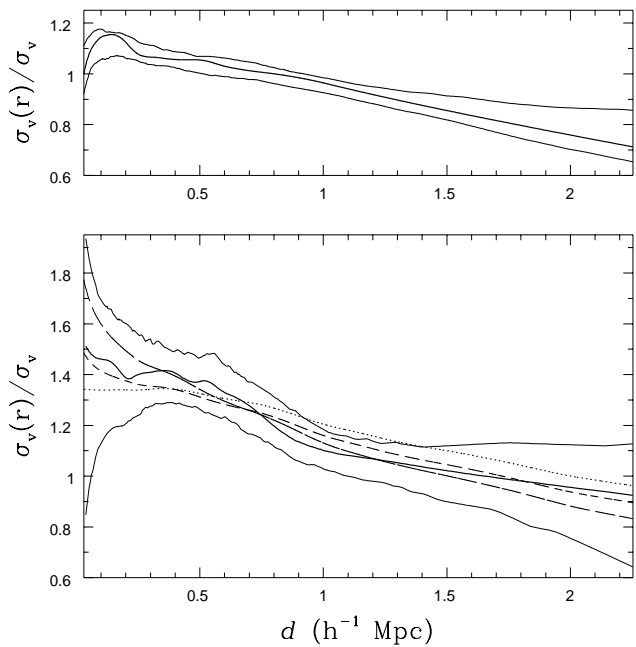,height=9.0cm,width=8.8cm}
\caption[]{The velocity-dispersion profiles for non-ELG (upper panel) 
and ELG (lower panel) in the synthetic cluster. The full-drawn curves
refer to the data, with heavy lines representing the observed profiles
while the thin lines indicate the 95\% confidence limits. In the lower
panel, three model predictions for different values of the anisotropy
parameter $\cal A$ are given, viz. $\cal A$ = -0.6 (dotted), 0.0
(short dashes) and 0.6(long dashes). }
\label{f-disprof}
\end{figure}

We think that the evidence for anisotropy of the ELG orbits is, in
fact, quite a bit stronger if one considers not just the inner and
outer values of \sv but includes the total velocity dispersion
profile. In Fig.~\ref{f-disprof} we show the observed velocity
dispersion profiles for the non-ELG (upper panel) and the ELG (solid
lines in the lower panel) in the synthetic cluster. The \sv profiles
were derived with the LOWESS method (Gebhardt et al. 1994). The heavy
lines represent the observed \sv, the two thin lines on either side
indicate the 95\% confidence bands, obtained from 1000 Monte Carlo
simulations of the dataset (for details, see e.g. Gebhardt et al.)
The increase of the uncertainty in the estimate of the ELG \sv for
r$_{proj} \la $ 0.3\Mpc is due to the flat ELG surface-density
profile. It must be realized that the values in Tab.~\ref{t-sigmas2}
are weighted averages over the indicated intervals, where the weights
follow from the observed surface densities shown in
Fig.~\ref{f-dprof}. In the lower panel we also show the predicted
velocity dispersion profiles of ELG for three models, viz. those with
$\cal A$ = -0.6, 0.0 and 0.6

On the basis of the steep gradient of the ELG \sv within 1.0 \Mpc, we
consider it quite unlikely that for the ELG the value of $\cal A$ is
practically 0.0 (i.e. that the ELG orbits are isotropic). Instead, the
sharp increase of the ELG \sv within 1.0 \Mpc is produced by galaxies
that mostly are at quite some distance from the centre where, in our
model, $\sigma_{rad}$ has already decreased considerably from its
central value $\sigma_{rad}$(0). To produce the observed values of the
line-of-sight velocity dispersion would seem to require inevitably
that the line-of-sight velocity component is significanty `boosted' by
radial anisotropy. Combining the data in Tab.~\ref{t-sigmas2} and the
information in Fig.~\ref{f-disprof}, we conclude that the best
estimate for $\cal A$ is probably $\approx$0.6$\pm$0.3 which implies a
value for $\sigma_{tan}/\sigma_{rad}$ between $\approx$0.3 and 0.8. Note
that we have not given much weight to the data within 0.3\Mpc due to
their large uncertainty. Had we given those more weight, we would have
arrived at a lower value of $\cal A$, probably between 0.0 and
0.3. However, for those values of $\cal A$ the predicted velocity
dispersion profile seems to give a rather bad representation of the
observed slope between 0.3 and 1.0\Mpc.

As mentioned before, the shape of the {\em full distribution} of
line-of-sight velocities (rather than only the dispersion) at
different r$_{proj}$ could, in principle, provide information on the
orbits. However, we have calculated those distributions for the models
summarized in Tab.~\ref{t-sigmas2} and found that with our statistics
the shapes are not a sufficiently sensitive discriminant.

\section{Discussion and conclusions}
\label{s-discuss}

Our study of the properties of emission-line galaxies in clusters has
yielded several results which we will now try and put together into a
more or less coherent picture. Some results are not unexpected and
confirm earlier results by other authors. On the other hand, we also
obtained some results that are totally new (among which the analysis
of the ELG orbits), and which are based on a sample of many tens of
rich clusters. Thereby our data provide evidence for the general
occurence of dynamical effects that up to now were seen only in one or
two individual clusters.

In the following discussion it must always be clearly realized that
our ELG simply are galaxies that had {\em detectable} emission lines
in the ENACS spectra. In several instances we will think of them as
(mostly late-type) spirals. This is justified as we found (for a
subset) that {\em well over 90\% of them were classified as spirals}.
However, they represent {\em only about 1/3 of the {\bf total}
spiral population}, as a result of our observational limit, and the
variation in gas content of spirals of different types. When comparing
the ELG with the non-ELG, we are thus always comparing a very
homogeneous class (ELG; read: spirals) with a heterogeneous class
consisting of both non-spirals (ellipticals and SO's) and spirals. 

The most striking results that we obtained concern the spatial
distribution and the kinematics of the ELG, especially when compared
with those of the non-ELG. 

First, the ELG very clearly avoid the central regions of clusters, and
the difference in central concentration of ELG and non-ELG probably
implies an even larger difference in the central concentration of
spirals and non-spirals. In the cluster Abell~576, Mohr et al. (1996)
recently also found a clear deficit of ELG in the central region. The
different spatial distributions are totally consistent with the
well-documented dependence of the mix of early- and late-type galaxies
on local galaxy density. The clear dependence of the fraction of ELG
on the global \sv of a cluster that we found can easily be understood
as a manifestation of this dependence. Several physical mechanisms
have been proposed for the dependence of galaxy mix on local density.
We will argue below that the kinematics of the ELG make it likely that
they still contain gas that produces detectable emission lines because
they have not yet been inside the high-density central cluster region.

The characteristic of the kinematics of the ELG that supports this
explanation most convincingly is their high velocity dispersion. For
the reasons explained above, we expect the observed ratio of the \sv's
of ELG and non-ELG of $\approx$1.2 to translate into a larger \sv
ratio for spirals and non-spirals. In this respect it is noteworthy
that Colless \& Dunn (1996) find that \sv of the late-type galaxies
in the main concentration in the Coma cluster is very close to $\sqrt
2$ times that of the early-type galaxies, which they interpret as
suggesting that the late-type galaxies are freely falling into the
cluster core.

Since we applied an interloper removal criterion irrespective of
whether a galaxy was classified as ELG or non-ELG, all our ELG
(including those projected onto the cluster core) are cluster members,
i.e. are within the present turn-around radius of their cluster. We
expect therefore that, had we been able to compare \sv of the spirals
with that of the early-type non-ELG (i.e. excluding the non-ELG
spirals) we would have obtained the same result as did Colless \&
Dunn.

Our result of a systematically larger \sv for ELG than for non-ELG
(which is based on an ensemble of many clusters) is supported by
recent observations of some individual clusters; Mohr et al. (ibid.)
find a similar effect in Abell~576, and Carlberg et al. (1996)
conclude for a sample of about 15 clusters with redshifts between 0.15
and 0.55 that `... the bluer galaxies, which often contain measurable
emission lines, statistically are found to have a higher velocity
dispersion than the redder absorption line galaxies, an effect that is
particularly prominent near the projected center of the cluster
...'. 

We believe that the larger \sv of ELG (and of the spirals) is a
generic aspect of the dynamics of galaxy clusters. It probably
indicates that the spirals that we see today avoid the central regions
because they either have not yet got there (the free-fall time is
certainly not much shorter than the Hubble time), or have passed by
the core on orbits that did not traverse the very dense central
region. In other words: the dynamical state of the ELG reflects the
phase of fairly ordered infall (of spirals) rather than the virialized
condition in the relaxed core, the size of which is probably only
$\approx$0.5 \Mpc (e.g. den Hartog and Katgert 1996). 

In this picture, the orbits of the ELG (and therefore of the spirals)
are expected to be fairly radial, and their velocity distribution is
expected to be quite anisotropic. The statistical weight of our
synthetic cluster with 549 ELG has allowed, we think for the first
time, a meaningful check of this prediction to be made. The
uncertainties of the ratio of the inner and outer \sv's still prevent
the anisotropy parameter ${\cal A}$ to be solved for with high
precision. However, the strong rise of \sv of the ELG towards the
centre, which was also seen by Mohr et al. in Abell~576 and by
Carlberg et al. (see above), quite strongly supports the notion of
predominantly radial orbits of at least the ELG that are projected
onto the central region.

A moderate to fairly strong anisotropy of the velocity distribution of
the ELG can also solve the apparent discrepancy of the mass estimates
based on ELG and non-ELG. We do not have a very accurate estimate of
the magnitude of the discrepancy because the non-ELG category does
contain spirals. Yet, the mass derived from spirals will (in terms of
the mass indicated by the non-spirals) be at least as large as that
derived from the ELG compared to the mass of the non-ELG, unless the
kinematics of the non-ELG spirals is totally different from that of
the ELG. From the discussion in \S~\ref{s-formev}, we estimate a
discrepancy of at least a factor 1.5. In the projected mass estimator
\begin{equation}
M_{PM} = \frac{f_{PM}}{GN} \sum_{i} v_{1d i}^{2} r_{\perp i}
\end{equation}
where $N$ is the number of galaxies in the system, $v_{1d i}$ the
observed velocity along the line-of-sight, and $r_{\perp i}$ the
projected clustercentric distance of the $i-th$ galaxy (Heisler et
al. 1985), the factor $f_{PM}$ is a projection factor that depends on
the distribution of orbits. It is equal to $64/\pi$ and $32/\pi$ for
the cases of radial (${\cal A}=1$) and isotropic (${\cal A}=0$)
orbits, respectively. More generally, one can show that
\begin{equation}
f_{PM} = \frac{4-2{\cal A}}{4-3{\cal A}} \frac{32}{\pi} = f({\cal A})
\frac{32}{\pi}
\end{equation}
where the anisotropy parameter ${\cal A}$ is assumed constant
throughout the system (see also Perea et al. 1990). If the ELG indeed
are not in equilibrium with the relaxed core, the mass estimate based
on them could be twice as large as the one based on the other
galaxies. This means that f(${\cal A}$) may be as large as 4/3, which
implies that ${\cal A}$ may be as large as 0.7, which is consistent
with the best value for ${\cal A}$ derived in \S~\ref{s-formev}.

The assumption that the spirals that we observe today are mostly
falling in for the first time is also consistent with the fact that,
contrary to earlier claims, we do not see any need for different
emission-line properties of ELG in clusters and ELG in the field. In
this respect it is noteworthy that only a fairly small fraction of the
ELG occur in compact subgroups. Our data are thus consistent with a
picture in which the infall of the spirals is rather isotropic.
Whether this is indeed so, or an artefact of our analysis, in which we
combined many clusters most of which contain only a fairly small
number of ELG, will become clear as soon as the results from the more
extensive multi-object cluster spectroscopy, that is presently under
way, will become available.

\begin{acknowledgements}

{It is a pleasure to thank Tim Beers, Gigi Fasano and Karl Gebhardt who
kindly provided us with Fortran codes for the evaluation of robust
statistics ({\em ROSTAT}), the 2-D KS-test, and the evaluation of
the velocity dispersion profiles ({\em LOWESS}), respectively, and 
Alan Dressler for kindly providing us with an electronic copy
of his data-set on cluster galaxy morphologies.

AB and PK thank Walter Jaffe for asking helpful questions about the
modelling. AM acknowledges financial support from the French GDR
Cosmologie and INSU; JP acknowledges support from the Spanish DGICYT
(program PB93-0159); and PF acknowledges support from the Italian
MURST.}

\end{acknowledgements}

\vfill
\end{document}